\documentclass{aa}
\usepackage{graphicx}
\usepackage{txfonts}

\def\l{$\lambda$}

\def\rfe{$R_{\rm FeII}$}
\def\Mb{$M_{\rm B}$\/}
\def\feiiq{\rm Fe{\sc ii}$\lambda$4570\/}

\def\msol{M$_\odot$\/}

\def\ltsima{$\; \buildrel < \over \sim \;$}
\def\simlt{\lower.5ex\hbox{\ltsima}}            % < over MMM
\def\gtsima{$\; \buildrel > \over \sim \;$}

\def\simgt{\lower.5ex\hbox{\gtsima}}            % > over MMM

\def\ha{{\sc H}$\alpha$}

\def\civ{{\sc{Civ}}$\lambda$1549\/}
\def\civnc{{\sc{Civ}}$\lambda$1549$_{\rm NC}$\/}
\def\civbc{{\sc{Civ}}$\lambda$1549$_{\rm BC}$\/}

\def\cm3{cm$^{-3}$\/}
\def\hb{{\sc{H}}$\beta$\/}

\def\hbbc{{\sc{H}}$\beta_{\rm BC}$\/}
\def\hbnc{{\sc{H}}$\beta_{\rm NC}$\/}
\def\mgii{{Mg\sc{ii}}$\lambda$2800\/}

\def\oiiiopt{{\sc{[Oiii]}}$\lambda\lambda$4959,5007\/}
\def\o4363{{\sc{[Oiii]}}$\lambda$4363\/}

\def\feiiopt{{Fe\sc{ii}}$_{\rm opt}$\/}
\def\feii{{Fe\sc{ii}}$_{\rm opt}$\/}
\def\fe{{\sc{Fe}}\/}

\def\heii{{He\sc{ii}}$\lambda$4686\/}
\def\fe76087{[Fe{\sc vii}]$\lambda$6087\/}
\def\oiii{{\sc [Oiii]}$\lambda$5007}

\def\kms{km~s$^{-1}$}
\def\rk{$R_{\rm K}$\/}
\def\ergss{ergs s$^{-1}$\/}
\begin{document}

\title{VLT/ISAAC Spectra  of the H$\beta$ Region in
Intermediate Redshift Quasars\thanks{Based on observations 
collected at the European
Southern Observatory, Chile. Proposal ref.: ESO 68.B--0364(A)}}

   \author{
            J. W. Sulentic \inst{1},
            G. M. Stirpe \inst{2}, P. Marziani\inst{3},
            R. Zamanov \inst{3,4},
            M. Calvani\inst{3},    \and
            V. Braito\inst{3,5}
          }

   \offprints{J. W. Sulentic}

\institute{   Department of Physics and Astronomy, University of
              Alabama, Tuscaloosa, AL 35487, USA\\
              \email{giacomo@merlot.astr.ua.edu}
\and
               Osservatorio Astronomico di Bologna, INAF,
               Via Ranzani 1, 40127   Bologna, Italy\\
               \email{giovanna.stirpe@bo.astro.it}
\and
      Osservatorio Astronomico di Padova, INAF,
              Vicolo dell' Osservatorio 5, 35122 Padova, Italy\\
              \email{marziani@pd.astro.it;
                     zamanov@pd.astro.it;
                     calvani@pd.astro.it}
\and
             Astrophysics Research Institute,
             Liverpool John Moores University,  Twelve Quays House,  Egerton Wharf,
             Birkenhead CH41 1LD,  United Kingdom\\
             \email{rz@astro.livjm.ac.uk}
\and
             Osservatorio Astronomico di Brera, INAF,
             Via Brera 28, 20121 Milano, Italy\\
             \email{braito@brera.mi.astro.it}
             }

\authorrunning{Sulentic et al.}
\titlerunning{VLT/ISAAC Observations of \hb\ in quasars}

\date{Received  / Accepted}

\abstract{
We present high S/N spectra of the H$\beta$ region
in 17 intermediate redshift ($0.85\leq$ z $\leq2.5$) quasars.
The spectra represent first results of our campaign to test the 
redshift/luminosity robustness of the so-called Eigenvector 1 
(E1) parameter space as developed for low redshift AGN in 
Sulentic et al.\ (\cite{S00b}). The new quasars span the luminosity 
range $-26\geq$~M$_B\geq -29$ while most of our low redshift sample
(n=215) involve sources in the range $-19\geq$~M$_B\geq -26$.
The high redshift sources show E1 parameter values and domain 
occupation that are similar to our low redshift sample 
supporting earlier findings that E1 parameters are uncorrelated 
with source luminosity. Elementary accretion theory can account 
for a systematic increase of the minimum observed H$\beta$ 
profile width with source luminosity. Narrow line Seyfert 1 
sources with M$_B= -28$ show FWHM(H$\beta$) as much as
2000~km~s$^{-1}$ broader than those with M$_B= -22$.  A possible
change in the higher redshift/luminosity sources involves
systematically weaker [OIII]$\lambda\lambda$4959,5007  narrow line
emission.

}

\maketitle

\section{Introduction}

There is as yet no convincing evidence for strong spectral evolution  
in quasars especially as far as low-ionization emission lines (LIL)
are concerned. Recent UV FeII observations suggests, for example, 
that FeII emission remains strong up to z$\approx$ 6.4 (Barth et al.\ \cite{barth};
Freudling et al.\ \cite{freudling}). The lack of spectral evolution may not present 
difficulties for quasar modeling {\it per se} but it may have serious 
cosmological implications (e.g. Hamann \& Ferland \cite{hamann}; Matteucci 
\& Recchi \cite{matteucci}). In the modelling context see Zamanov \& Marziani (\cite{ZM02})
for a demonstration of self-similar properties in widely different 
accreting systems. 

We have been searching for a parameter space to serve the role of the stellar
H-R Diagram in discriminating quasar spectral phenomenology and evolutionary 
states. Our so-called Eigenvector 1 (E1) parameter space (Sulentic et
al.\ \cite{S00b}, hereafter S00b) shows promise in this context (see e.g.
Marziani et al.\ \cite{M01}; Sulentic et al.\ \cite{S02}; Marziani et al.\ \cite{M03a}).
The optical E1 parameters involve broad line measures of the full
width at half-maximum (FWHM) of the broad component of H$\beta$
(\hbbc) and the equivalent width ratio \hbbc/FeII, where FeII
is measured from the $\lambda$4570 blend. These are supplemented by
measures of higher ionization CIV$\lambda$1549 line shift and the soft
X-ray photon index making E1 a 4D parameter space. The distribution of
sources in the E1 optical plane is consistent with a principal band
or ``main sequence'' of source occupation. The shape of that principal
sequence motivated an alternative suggestion that two AGN populations
exist with an arbitrary separation at FWHM \hbbc=4000~km~s$^{-1}$.
Population A sources generally show radio-quietness, strong optical
FeII emission, a soft X-ray excess and a systematic CIV blueshift.
Narrow line Seyfert 1 sources (NLSy1) represent an extreme of
Population A which contains $\sim$65\% of all radio-quiet (RQ) sources
(Marziani et al.\ \cite{M03a}). Population B sources generally show weaker
FeII emission and no soft  X-ray excess or CIV blueshift. The latter
population contains most radio-loud (RL) sources, and about 25\% of
the RQ sources. RL sources found in Population A are located there
because of a preferred orientation to our line of sight (e.g.
core-dominated radio emission; Sulentic et al.\ \cite{S03}). These properties
have emerged from a growing sample (N$\geq$215; Marziani et al.\ \cite{M03b},
hereafter M03b) of low redshift (mostly z$<$0.8) AGN. Results so-far
give us cautious optimism that source orientation can be decoupled
from physics using E1 space (Marziani et al.\ \cite{M01}; Sulentic et al.\
\cite{S03}).

Our definition of E1 began with the low redshift part of the 
PG quasar sample (Boroson \& Green \cite{BG92}) and our optical 
E1 parameters emerged in their principal component analysis of the
correlation matrix for the PG sample. Interestingly enough, source luminosity
emerged in their second eigenvector implying that the E1
correlations are source luminosity independent at least at low z. We 
continue to find no evidence for correlations involving optical 
luminosity in E1 space. Radio luminosity is correlated but {\it only} 
in the sense that radio-loud AGN show significantly restricted domain 
space occupation (e.g. Population B). At the same time, sources with
radio/optical flux ratio $< 10$ (RQ) show no preferred domain space 
occupation in E1.

Naturally we would like to test the robustness 
of the E1 space using samples of sources with higher mean 
redshift and luminosity. This involves tests using samples 
of quasars with z~$>1.0$ and optical spectroscopy CIV and/or
IR measures of the H$\beta$ region. 
Recent Sloan Digital Sky Survey (SDSS) results for a large 
intermediate/high redshift sample 
(Richards et al.\ \cite{richards}) show CIV trends very similar to our low
redshift (S00b) E1 results (i.e. 65--75\% of sources
with Population A CIV properties). IR spectra of the H$\beta$ region, 
up to this point in time (e.g. Espey et al.\ \cite{espey}; Murayama et al.\ \cite{murayama}; 
Oya et al.\ \cite{oya}; McIntosh et al.\ \cite{Mc99}, hereafter Mc99; but see
Dietrich et al.\ \cite{dietrich}) have suffered from low resolution and S/N
making comparisons with our E1 sample impossible. We have begun a
campaign to obtain high S/N IR spectra of the region of the
H$\beta$+optical FeII$\lambda$4570 blend in intermediate redshift
quasars.
We present results from the reduction and analysis of ESO VLT1/ISAAC
spectra for our first year sample of 17 sources. We show that the
quality of these data is comparable to our low-redshift database
(M03b). We are able to measure the \hbbc\ and \feii\ 
E1 parameters using the same techniques described in M03b (Sect.~\ref{obs}). 
We present an analysis of the luminosity and redshift
trends using the new intermediate redshift IR and older low redshift
optical data (Sect.~\ref{lum}). We also interpret the new line measures
in the context of the  E1 parameter space (Sect.~\ref{e1}).

\section{Observations, Data Reduction, and Measurements\label{obs}}

Table \ref{tab:obs} presents a summary of the observations and basic source
properties, with footnotes giving detailed descriptions of each column. Data
were collected using VLT1/ISAAC operated in service mode between 2001
November and 2002 February. All spectra were obtained with a slit
width of 0.6~arcsec. Each spectrum corresponds to the wavelength range
of an IR window (sZ, Z, J, H) and covers the region of redshifted
H$\beta$ and \feiiq\ (or Fe{\sc ii}$\lambda$5130 blend). Two matching
spectra in adjacent bands were obtained in five cases, to improve the
coverage of the \hb\ spectral region.

Reductions were performed using standard IRAF routines.
Sequences of frames with a given Detector Integration
Time (DIT, see Table 1) were obtained with
the source at different positions (e.g. A, B, C...) along the slit.
All frames at a specific slit position were averaged
and the average of observations at all other positions
was subtracted from it.
The resulting differences were divided by the appropriate
flat field  frames provided by the ESO pipeline reduction.
Any residual background was then  eliminated by fitting and
subtracting a low-order polynomial function to each
spatial line of the frame. Spectra were extracted using the
IRAF program {\it apall.} Cosmic ray hits were
eliminated by interpolation, comparing the affected spectrum with the
other spectra of the same source. For each position along the slit a
corresponding Xenon/Argon arc spectrum was extracted from the
calibration frame, using the same extraction parameters.
The wavelength calibration was well modeled by 3$^{\rm rd}$ order
Chebyshev polynomial fits to the positions of
15--30 lines, with rms residuals of 0.3~\AA\ in the Z band, 0.4~\AA\
in sZ, 0.6~\AA\ in J, and 0.9~\AA\ in sH. Once matched with the
corresponding arc calibrations, the individual spectra of each source
were rebinned to a common wavelength scale. They were then averaged
with weights proportional to the total integration time of each
spectrum.

The spectra of the atmospheric standard stars were extracted and
wavelength-calibrated in the same way. All clearly identifiable stellar
features (H and HeI absorption lines) were eliminated from the stellar spectra
by spline interpolation of the surrounding continuum intervals. Each target
spectrum was then divided by its corresponding standard star spectrum in order
to correct for the atmospheric absorption features. This was achieved with the
IRAF routine {\it telluric,} which allows one to optimize the correction with
slight adjustments in shift and scaling of the standard spectrum. The shape of
the continuum of the standard star was eliminated from the spectrum of each
target by multiplying the latter with an artificial black-body continuum
corresponding to the temperature of the star, determined on the basis of its
tabulated spectral type. Finally, the correct flux calibration of each
spectrum was achieved by scaling it according to the magnitude of the standard
star and to the ratio of the respective DITs. Because the seeing
almost always exceeded the width of the slit, significant light loss
occurred, and therefore the absolute flux scale of the spectra is
not to be considered as accurate. However, in this high-wavelength
range we consider the light losses to be independent of wavelength,
and they should therefore not affect the relative calibration of the
spectra.

\begin{figure*}
  \includegraphics[width=9.6cm, height=13.6cm, angle=0]{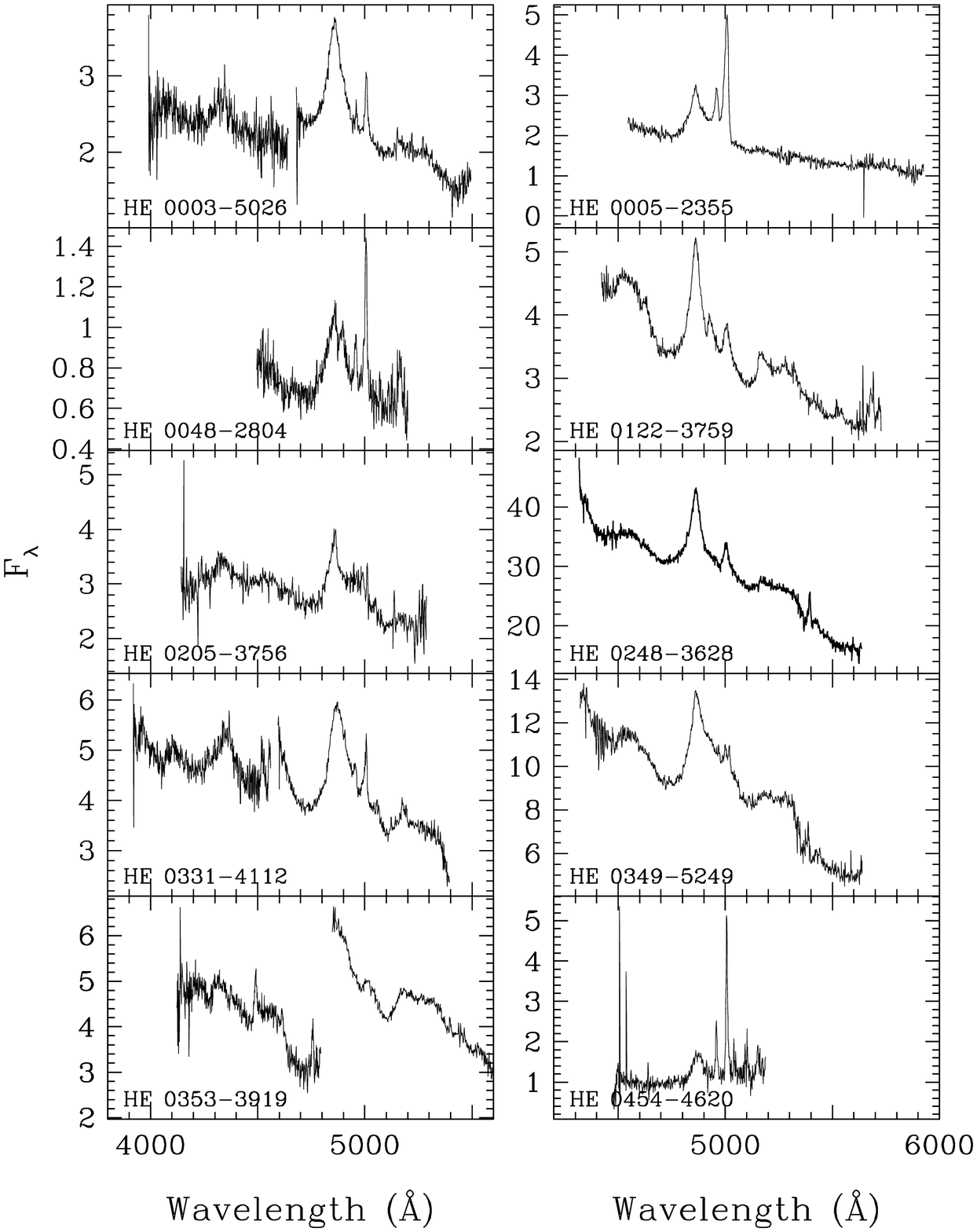}
  \includegraphics[width=9.6cm, height=13.6cm, angle=0]{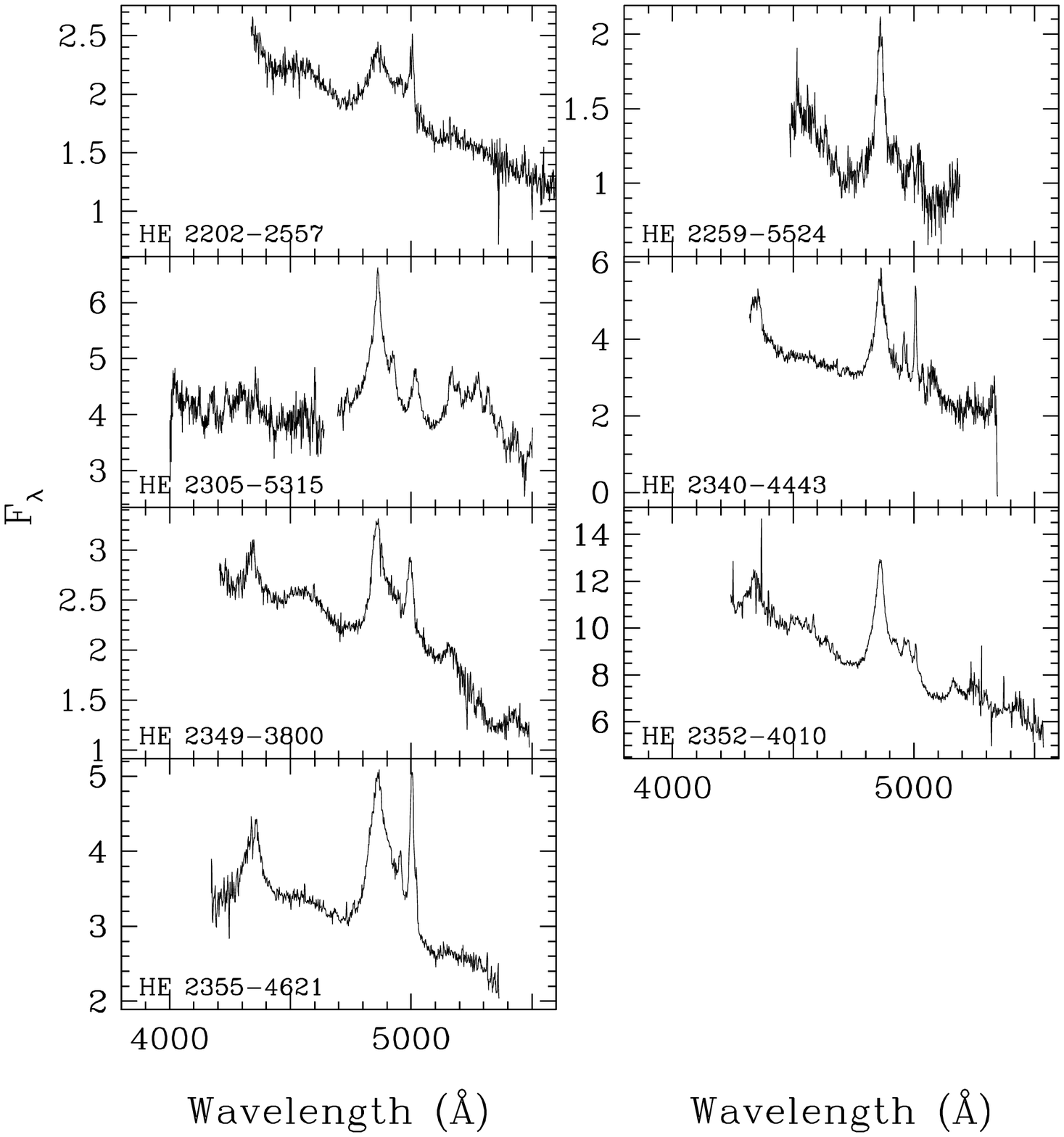}

      %\vspace{22cm}
      \caption[]{Calibrated spectra of the 17
intermediate-redshift quasars. Abscissae are rest-frame wavelength in \AA,
ordinates are specific flux in units of 10$^{-15}$ \ergss\ cm$^{-1}$
\AA$^{-1}$. Note that, for \object{HE 0353$-$3919}, $\sim 100$~\AA\ are missing
in the blue side of the \hbbc\ profile, due to a gap between the sZ
and Z bands. }          \label{fig:spectra}
   \end{figure*}

\begin{figure*}
  \includegraphics[width=9.6cm, height=14.6cm, angle=0]{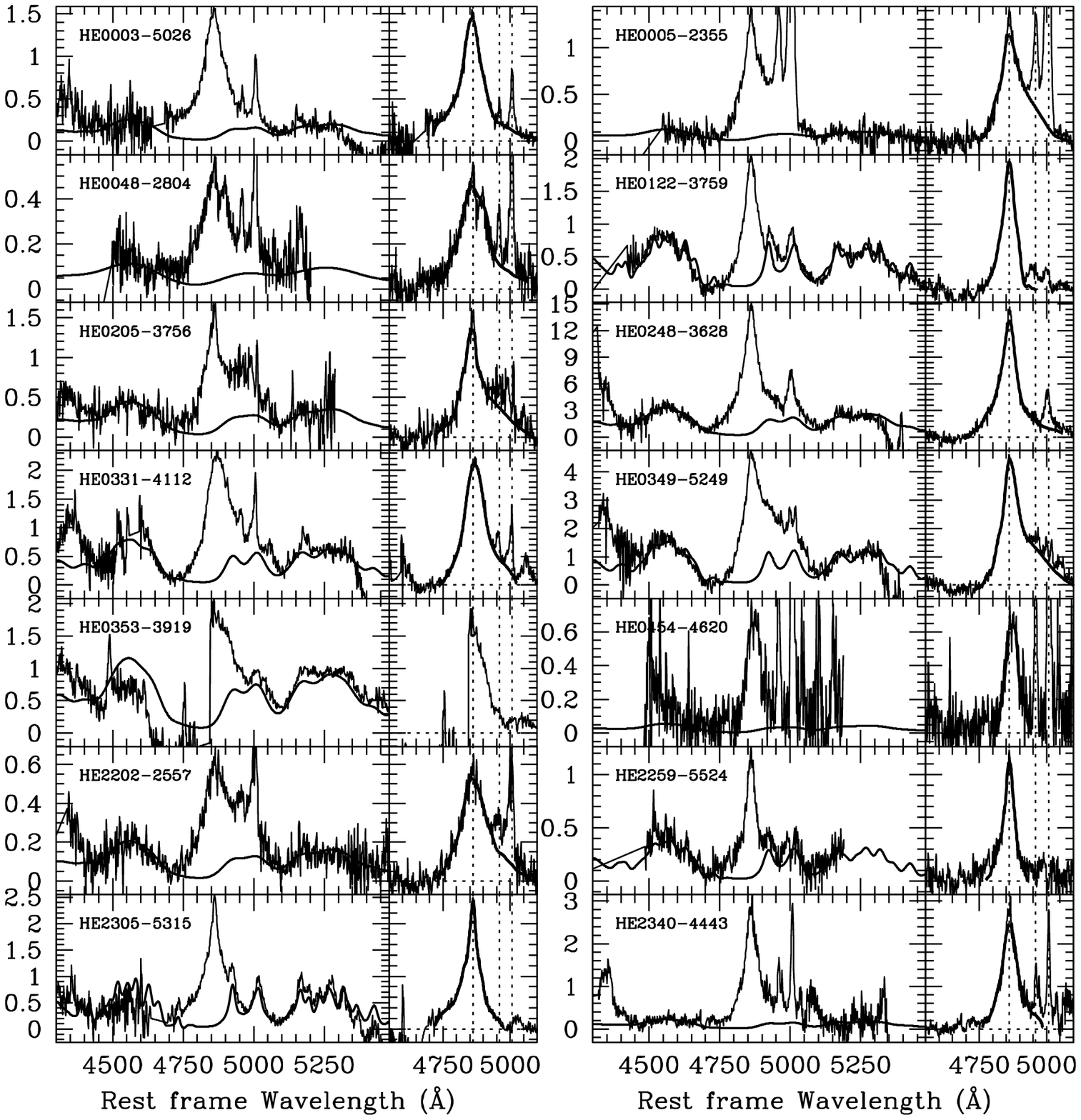}
  \includegraphics[width=9.6cm, height=14.6cm, angle=0]{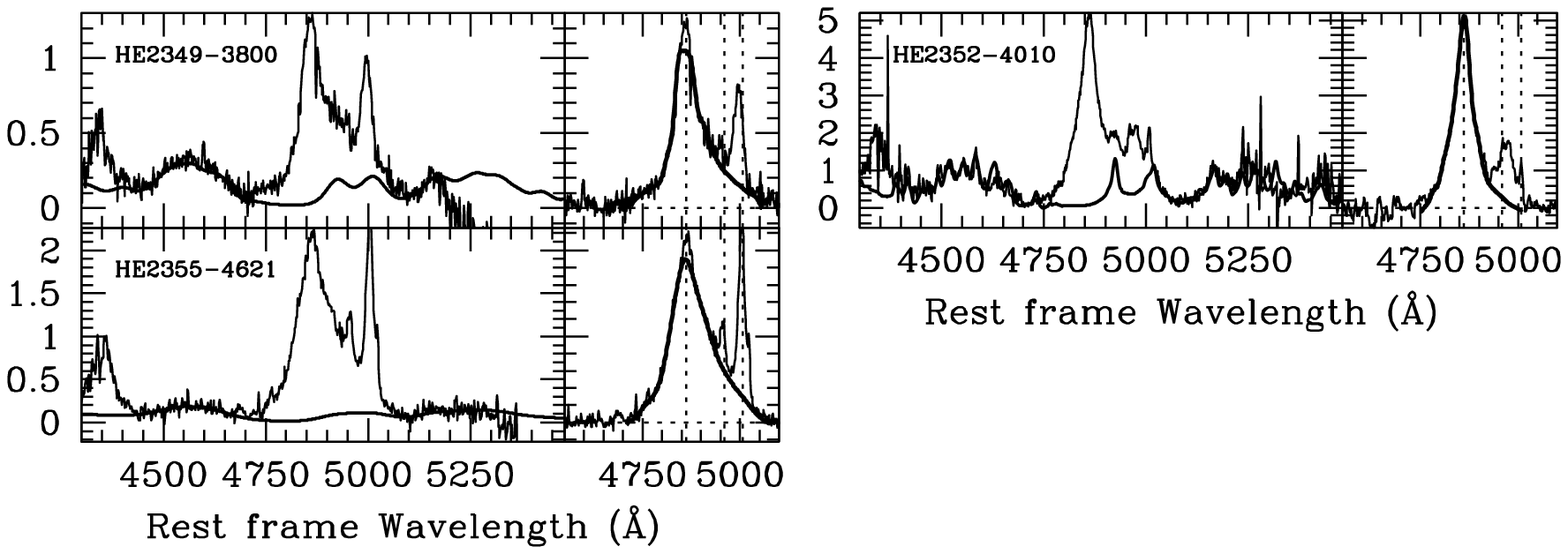}

      %\vspace{22cm}
      \caption[]{Spectral Atlas of the 17
intermediate-redshift quasars. The left panels show the continuum-subtracted \hb\
spectral region. Abscissae are rest frame wavelength in \AA, ordinates are
specific flux in units 10$^{-15}$ \ergss cm$^{-1}$ \AA$^{-1}$. The
best-fit \feii\ emission (see text) is traced as a thin (green) line.
The right panels show an expansion around \hb\ of the same spectrum
after continuum and \feii\ subtraction. Abscissae and ordinates are as
above. The (blue and red) thick line shows a spline fitting of the
pure \hbbc\ on the short and long wavelength side of the line
respectively. }          \label{fig:atlas}
   \end{figure*}

Redshift measures are usually
based on \oiii\ but the narrow component of H$\beta$ (\hbnc) was also
used whenever possible. Redshift uncertainty is usually $\la150$~\kms,
including estimated uncertainty of the wavelength calibration. These
measures can be regarded as the most accurate available, with the
caveat that some (Population A) sources with narrowest FWHM \hbbc\
sometimes show a significant \oiiiopt\ blueshift (Zamanov et al.\
\cite{Z02}; Marziani et al.\cite{M03a}). The S/N ratio has been estimated by: (1)
locating a spectral region that is flat and free of strong lines and
(2) dividing two times the rms scatter by the average signal in that
region. Examination of the spectra will reveal the limited regions
available for S/N and continuum estimation. S/N values are comparable
to our low-redshift M03b data. The IR spectra presented in Figs.\
\ref{fig:spectra} and \ref{fig:atlas} are, on average,
indistinguishable from the data in the M03b atlas. Spectral resolution
is FWHM~=~9~\AA\ in the Z band ($\lambda_{\rm C} \approx 9000$~\AA),
11~\AA\ in the sZ band ($\lambda_{\rm C}\approx 10600$~\AA), 12~\AA\
in the J band ($\lambda_{\rm C}\approx 12400$~\AA), and 16~\AA\ in the
H band ($\lambda_{\rm C} \approx 16000$~\AA). In all bands this is
equivalent to FWHM~$\approx$ 300~\kms\ which is similar to the
resolution of the M03b data.

Measurements were carried out with exactly the same technique
employed by M03b. The de-redshifted spectra were
continuum- and then \feii-subtracted. The spectral
width of the IR windows made continuum modeling and subtraction
uncertain in many cases. The lowest regions in the
adopted FeII fits shown in the spectra of Figure \ref{fig:atlas} allow one
to infer the adopted continuum level.  \feii\ subtraction was achieved
employing a template based upon spectra of \object{I~Zw~1}, scaled and broadened
by fixed factors in a plausible range chosen by
eye. The best \feiiq\ template was chosen as the one yielding the
minimum residual in a matrix of 10$\times$10 scaling and broadening
factors. An interesting result comes from the broadening factor of the
\feii\ template: an estimate of the intrinsic width of the individual
\feii\ lines. All measures  have an accuracy (for a given S/N)
similar to those in M03b. The \hbbc\
was isolated after subtracting the narrow component of \hbnc\
(self-consistent guidelines are provided in M03b). Both
\oiiiopt\ and \heii\ were also measured whenever possible.

\section{Sample Considerations\label{sample}}

We adopted the Hamburg-ESO (HE) quasar surveys (Wisotzki et al.
\cite{wisotzki}) for tests of E1 validity and robustness.
The $U-B$\ color-selected PG sample (Boroson \& Green \cite{BG92}) is
thought to be biased towards selecting what we call extreme
Population A sources i.e. NLSy1s (FWHM(\hbbc)~$\la2000$~\kms). We are
in the process of observing both  low and intermediate redshift
subsamples of grism-selected HE quasars in order to evaluate effects
of selection bias on mean E1 parameter values and E1 domain space
occupation. The low redshift sample will compare E1 properties 
of HE grism- vs. PG color-selected  quasars. The intermediate 
redshift sample will explore possible E1 changes  with 
redshift/luminosity. If anything, the HE samples should be biased 
towards broader/stronger lined (Population B) quasars. The PG 
sample finds 20/87 $\approx$ 23\%\ NLSy1s while the (also 
color-selected) SDSS (Williams et al.\ \cite{williams}) suggests that 
$\sim$~15\% of all low-redshift AGN are NLSy1. 
Our low redshift M03b sample includes 150 RQ and 65 
RL sources. NLSy1, which are very  rarely RL, account for 11\% 
and 16\%, respectively, of the total  and RQ parts of our sample. 

RL sources are over-represented (30\%) in our sample because that
part of our low-redshift sample has been surveyed more completely
to our adopted magnitude limit $V=16.5$. We find 85\% RQ and
10\% RL in the Population A domain
while 37\%\ RQ and 75\%\  RL are found in the Population B domain.
Only 7\%\ of the sample  fall off of the Population A-B ``main
sequence'' and are designated ``outliers''. RL sources are defined as
those with $R_{\rm K} = f_{\rm 6cm}$/$f_{\rm B}\ga 100$, plus
any sources near and below that
limit showing double-lobed (FRII) radio morphology (\rk~$\approx
70-80$; Sulentic et al.\ \cite{S03}).  All true core-dominated RL sources
are assumed to be radio flux-boosted FRII's and should show values of
\rk~$>80$.  The condition \rk~$\ga80$ yields 4 RL sources in our
VLT/ISAAC sample with two additional radio-intermediate sources with
$10 \la R_{\rm K} \la 80$. This implies an excess of RL sources in the
intermediate-redshift sample; however it is premature to draw such
conclusions. The small size of our new sample suggests that sources
are best compared in terms of the so-called E1 Populations A and B.
They are represented by 6 and 11 sources respectively.

\begin{figure*}
  \includegraphics[width=7.6cm, height=17.6cm, angle=270]{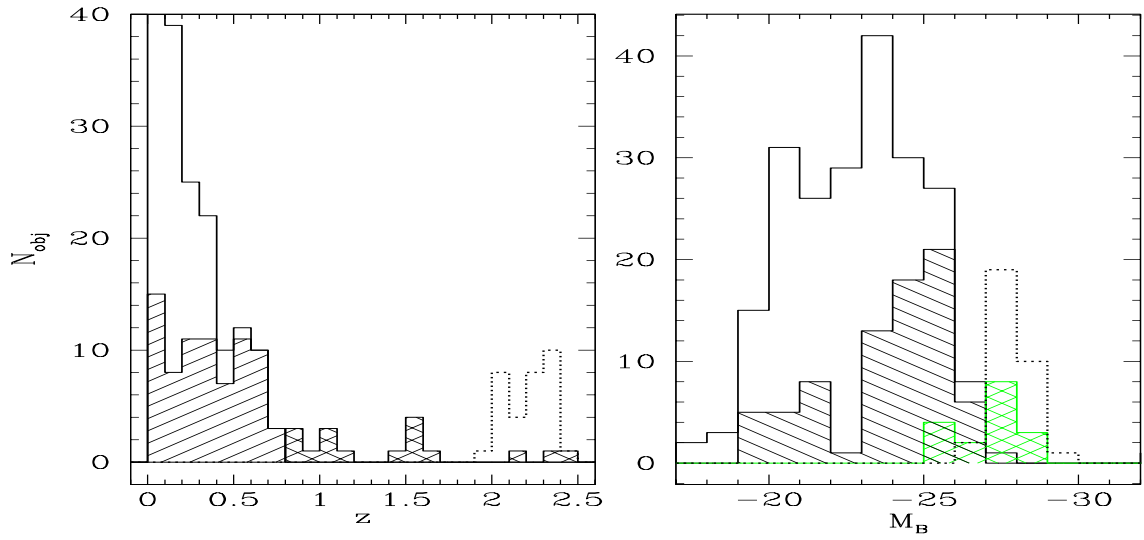}
      %\vspace{22cm}
      \caption[]{Redshift (left) and absolute B magnitude (right) distributions for the total M03 sample (unhatched),
      for the RL subsample of M03 (hatched), for the 17 sources of this study (cross-hatched) and for the sources
      with $2.0\la z\la 2.5$ studied by Mc99 (dotted line).}
         \label{fig:distr}
   \end{figure*}

Our comparisons with the low-$z$-defined E1, and search for luminosity
effects, make use of the M03b dataset that includes 215 sources  with a
``core'' of $\approx$~85 sources from the PG survey (Boroson \& Green
\cite{BG92}). In Marziani et al. (\cite{M03a}) we increased our low-redshift sample by
adding the soft-X-ray selected sources from Grupe et al.\ \cite{grupe}. All
defining properties of E1 space have remained stable
as our sample has grown from 70 to $>250$ low-redshift sources.
Strong luminosity effects were ruled out but luminosity
dependence was not studied in detail.  No claim of completeness can be
made for the majority RQ part of the M03b sample ($\sim$30\%), however
V/V$\rm _{\rm max}$ tests suggest that the RL part is about 
80\% complete to $z\approx0.8$; $m_{\rm V} \approx 16.5$.
However incompleteness is  not a major impediment to a proper 
correlation analysis with $M_{\rm B}$ if we have uniform sampling 
across the entire absolute magnitude range. Figure~\ref{fig:distr}
shows the \Mb\ (V\'eron-Cetty 
\& V\'eron \cite{veron}) and redshift distributions for the M03b and VLT
samples. We also include the much lower S/N high-$z$ observations from
Mc99.  The range $-20 \ga$~\Mb~$\ga -29$ is reasonably well sampled.
It is important to stress that, if RQ and RL sources are considered
separately, most RQ fall in the range  $-20 \ga$~\Mb~$\ga -28$,
while RL are on average more luminous, $-24 \ga$~\Mb~$\ga -29$.
We are observing the brightest sources in the HE survey which means that
we are sampling an $M_{\rm B}$\ range similar to Mc99 but with sources
distributed over a wider range of redshift.

\section{Results\label{res}}

Wavelength- and flux-calibrated  spectra of the  17 HE
quasars are shown in Figure \ref{fig:spectra} and Figure
\ref{fig:atlas}. Figures show de-redshifted spectra both
before and after continuum and \feii\ subtraction. The right-hand
panels of Fig. \ref{fig:atlas} show the ``cleaned" \hbbc\ profile
following  \hbnc, \oiiiopt\ and \heii\ subtraction. Rest-frame
equivalent widths are given in Figure \ref{tab:meas} for \hbbc, \feiiq,
\hbnc\ and \oiiiopt, along with the FWHM estimates for
individual terms of the \feiiq\ emission.  FWHM and other \hbbc\ 
profile measures are provided in Table \ref{tab:profs} 
along with 2$\sigma$ uncertainties. Line parameters
such as asymmetry index, kurtosis and line centroid at various
fractional intensities are the same as defined in Marziani et al.\ \cite{M96}
(hereafter M96) and M03b. Asymmetry index is defined as:
$${\rm A.I.(\frac{1}{4})}= \frac{v_{\rm R}(\frac{1}{4}) + v_{\rm B}(\frac{1}{4})
- 2 c(9/10)}{v_{\rm R}(\frac{1}{4}) - v_{\rm B}(\frac{1}{4})}$$
where $v_{\rm B}(\frac{1}{4})$\ and $v_{\rm
R}(\frac{1}{4})$ are the radial velocities measured on the blue and red sides
of \hbbc\ at 1/4 fractional intensity. The asymmetry index
thus defined is independent of assumptions about the rest frame. In
general, we define a centroid radial velocity as
$$c(\frac{i}{4})=\frac{v_{\rm R}(\frac{i}{4}) + v_{\rm B}(\frac{i}{4})}{2},$$
which we list in Table \ref{tab:profs} for $i=1$ and 3. In the
definition of A.I., we use $c(\frac{9}{10})$ as a proxy for the peak
radial velocity $c(\frac{4}{4})$.

\begin{figure*}
  \includegraphics[width=17.6cm, height=17.6cm, angle=0]{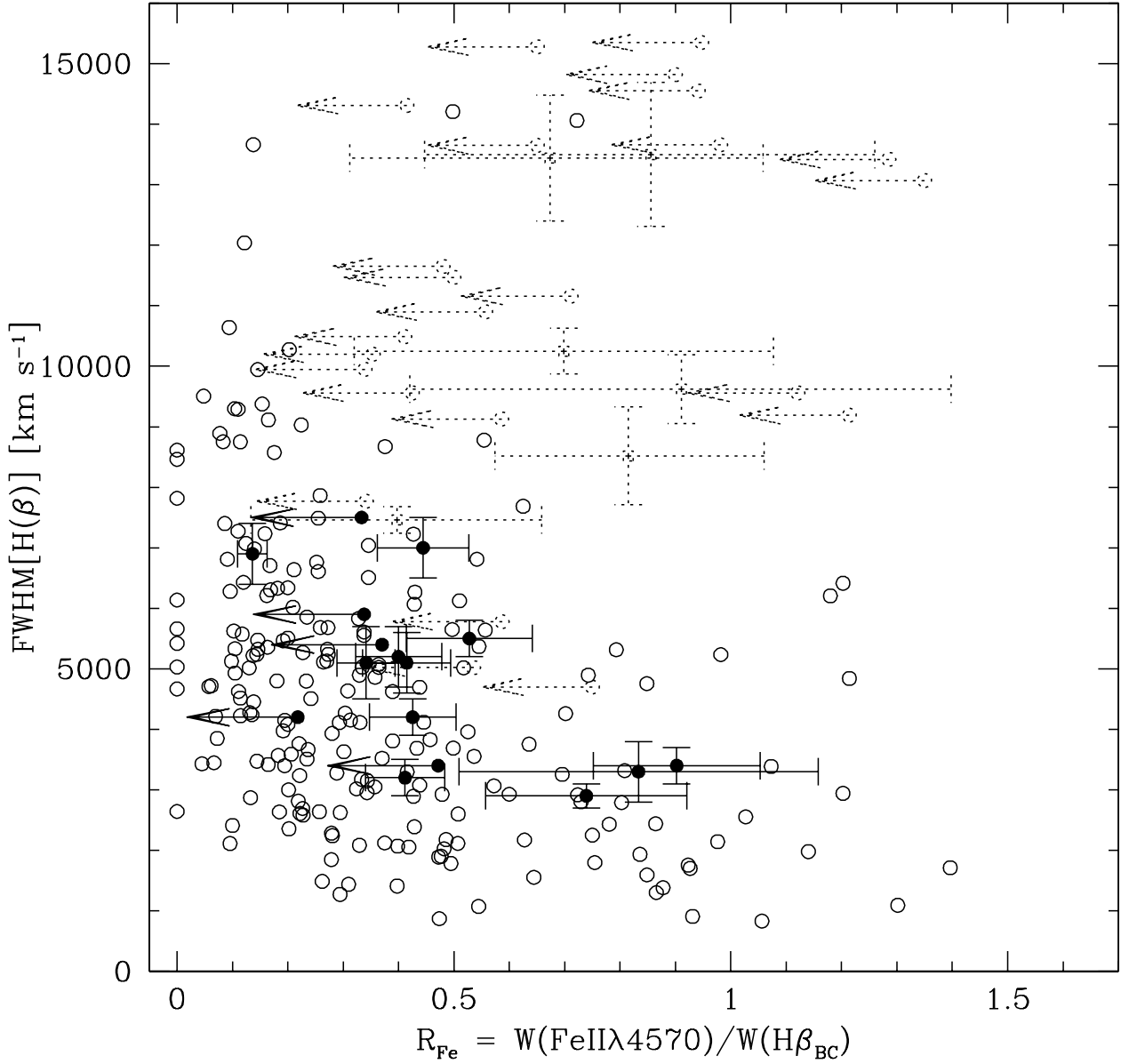}
      %\vspace{22cm}
      \caption[]{Distribution of Seyfert 1  and quasars in the optical plane
      of E1, FWHM(\hbbc) (in units of \kms) vs. \rfe. Open circles: M03, 215
      AGN; filled circles: 17 VLT/ISAAC sources of the present study; open circles
      with dotted lines: 22 $z \approx 2.5$ sources studied by Mc99, with upper limits and uncertainties
      in \rfe\ set according to M03. }
         \label{fig:e1}
   \end{figure*}

The optical Eigenvector 1 parameters (\rfe = W(\feiiq)/W(\hbbc) and
FWHM(\hbbc); S00b) can be computed from  the data 
in Tables \ref{tab:meas} and \ref{tab:profs}.  Figure
\ref{fig:e1}  shows the location of the low and intermediate-redshift
quasars in the optical plane of E1. The VLT/ISAAC sources show no
significant  difference in E1 domain space occupation.  In contrast 
Mc99 data suggests a rather different picture with most
sources located in an ``outlier'' region that is scarcely populated 
by low-redshift AGN: FWHM(\hbbc)~$\ga 10000$~\kms. No low-redshift
sources are observed
with the additional condition \rfe~$\ga 0.5$, where $\approx 10$ of
the Mc99 sources are found. We suggest that the latter data sample is
critically affected by poor S/N (\S \ref{mc}).

Some general trends  seen in the low-redshift data continue to be 
found: 1) RL sources favor Population B and show lower average \rfe\
values, 2) FWHM \hbbc\ and \rfe\ values do not differ significantly
between Population B RL and RQ sources, 3) moving towards Population
A we find the same systematic increase in average \rfe\ (from 0.37 to
0.54), 4) Population A sources favor the upper envelope of the low 
redshift distribution which may be pointing to a correlation between
FWHM(\hbbc) and luminosity (see 5.2.1) and 5) \civ\ measures show
expected E1 trends as discussed in the next section.

\subsection{CIV$\lambda$1549 Trends for the VLT Quasars}

CIV profile shift was adopted as one of the E1 parameters rather than 
EW CIV because its interpretation is less ambiguous. Population A sources 
in the low-redshift sample show a systematic CIV blueshift while
Population B sources do not. Population A sources also show a lower
mean EW than Population B AGN (S00b). Optical ground-based
\civ\ spectra exist for three of the HE sources and they are discussed
individually. In addition HE discovery spectra (courtesy of L.~Wisotzki)
for sources with $z\ga  1.5$ include \civ. Table \ref{tab:civ} 
summarizes the E1 CIV shift parameter for VLT sources with available data. 
Profile shift was measured  relative to rest frame measures derived 
from  \oiiiopt

\paragraph{\object{HE 0005$-$2355}:} We call this a Population B RL source.
Espey et al.\ (\cite{espey}) report $z\approx 1.411$ consistent with
the general absence of large \civ\  blueshifts in Population B sources
(we obtain $z\approx1.412$; Espey et al.\ \cite{espey} report $z\approx1.407$
for \ha\ with FWHM(\hbbc)~$\approx5900$~\kms). The more recent HE
measure gives $z\approx1.405$ but with CIV at the noisy blue edge of
the spectrum.

\paragraph{\object{HE 0122$-$3759}:} Population A RQ source. Comparison of
\civ\ redshifts (2.173: Carswell et al.\ \cite{carswell}; 2.178:  Espey et
al.\ \cite{espey}) with our \oiiiopt\ value (2.200) suggest a large blueshift.
Espey et al.\ (\cite{espey}) derive $z=2.207$ from \ha\ and $z\approx2.199$ from
\mgii. An HE spectrum (CIV $z\approx2.164$) confirms the large
blueshift $C\sim -3400$~\kms\ with an amplitude seen only in
extreme Population A (i.e. NLSy1) sources. The highest amplitude
blueshifts at high and low redshift fall in the
range $C \approx -4000\div-5000$~km/s (S00b; Richards et al.\
\cite{richards}).

\paragraph{\object{HE 0205$-$3756}:} Population A RQ source. A published
\civ\ measure yields $z\approx2.395$ (Ulrich \cite{ulrich}). This implies a
large blueshift $C\sim3000$~\kms\ relative to our rest frame
measure $z=2.437$ (2.412 in Wilkes \cite{wilkes86}). The low measured W(\civ)
(Ulrich \cite{ulrich}) is also consistent with a Population A source
(S00b). Taken at face value, EW and profile shape
for \hbbc\ appear characteristic of Population B. However the
\hbbc\ red shelf may be a spurious feature caused by 
residuals from the very strong sky lines. Either this object
is affected by bad data or it may herald a change towards 
more ``population ambiguous'' quasars at high redshift/luminosity 
that must be monitored as our sample increases. \\

In summary we confirm that  \civ\ blueshifts first observed 
by Gaskell (\cite{gaskell}) may be increasingly common in intermediate 
redshift quasars. All certain  blueshifts in the VLT HE sample  
are found in Population A sources as predicted from E1, while
all Population B shifts are marginal (see Table \ref{tab:civ}).

\subsection{Luminosity Trends\label{lum}}

Studies of the E1 parameter space have thus far been constrained 
to sources with $z < 1.0$ and mostly \Mb~$\ga -25.0$.
The  redshift constraint reflects the  loss of  the H$\beta$ spectral 
region to optical observation at $z\ga$~1.0.
The magnitude constraint reflects our S/N and resolution requirements 
convolved with telescopes readily available to us. Lower-quality data 
cannot provide accurate E1 parameter measures or  reveal E1 domain 
occupation clearly. The analogy would be to try to define 
the H-R diagram for a star cluster using magnitudes and 
$B-V$\ colors with respective uncertainties of $\pm$1.0 and $\pm$0.4.   
This issue would be irrelevant if all quasars were alike but source
occupation in the  E1 domain (in analogy to  stars in the H-R domain) 
is not random and the difference between so-called Population A and B  
quasars is found in virtually all AGN properties (see also Sulentic 
et al.\ \cite{S02}).  Within our sample  constraints, optical luminosity is
uncorrelated with  E1 properties  at low redshift (S00b; M03b).

Beyond tests of the robustness of E1 space, extension of our
sample to higher redshift/luminosity can address many questions.
Do quasars maintain the same emission line properties over the 
full range of redshift/luminosity? Can we constrain any form of evolution? 
In particular, can we identify any difference in 
optical \feii\ emission? Is the evolution of quasar spectral 
parameters consistent with the expectations of broadening by 
virial motions? We will address these questions, as far as currently
possible, one parameter at a time.
Table \ref{tab:corr} provides a synopsis of our luminosity correlation
analysis. We report the generalized Spearman rank
correlation coefficient computed for the general case of censored data
(Isobe et al.\ \cite{isobe}). M03b data yield only meaningful upper limits for 
\feiiq\ and \rfe\ for several tens of sources. Of course,
whenever upper limits are not considered the correlation coefficient
reduces to the usual Spearman $r$. We also considered  
the PG quasar sample independently as well as the
 VLT/ISAAC data and, with a single exception (see Sect.~\ref{e1lum}), found
behavior in agreement with those of the other samples (albeit the
PG RL sub-sample ($n=15$) is too small to give reliable correlation
coefficients). No credible evidence for a luminosity correlation 
was found. 

\subsubsection{E1 Broad Line Parameters\label{e1lum}}

\begin{figure*}
  \includegraphics[width=8.0cm, height=8.0cm, angle=0]{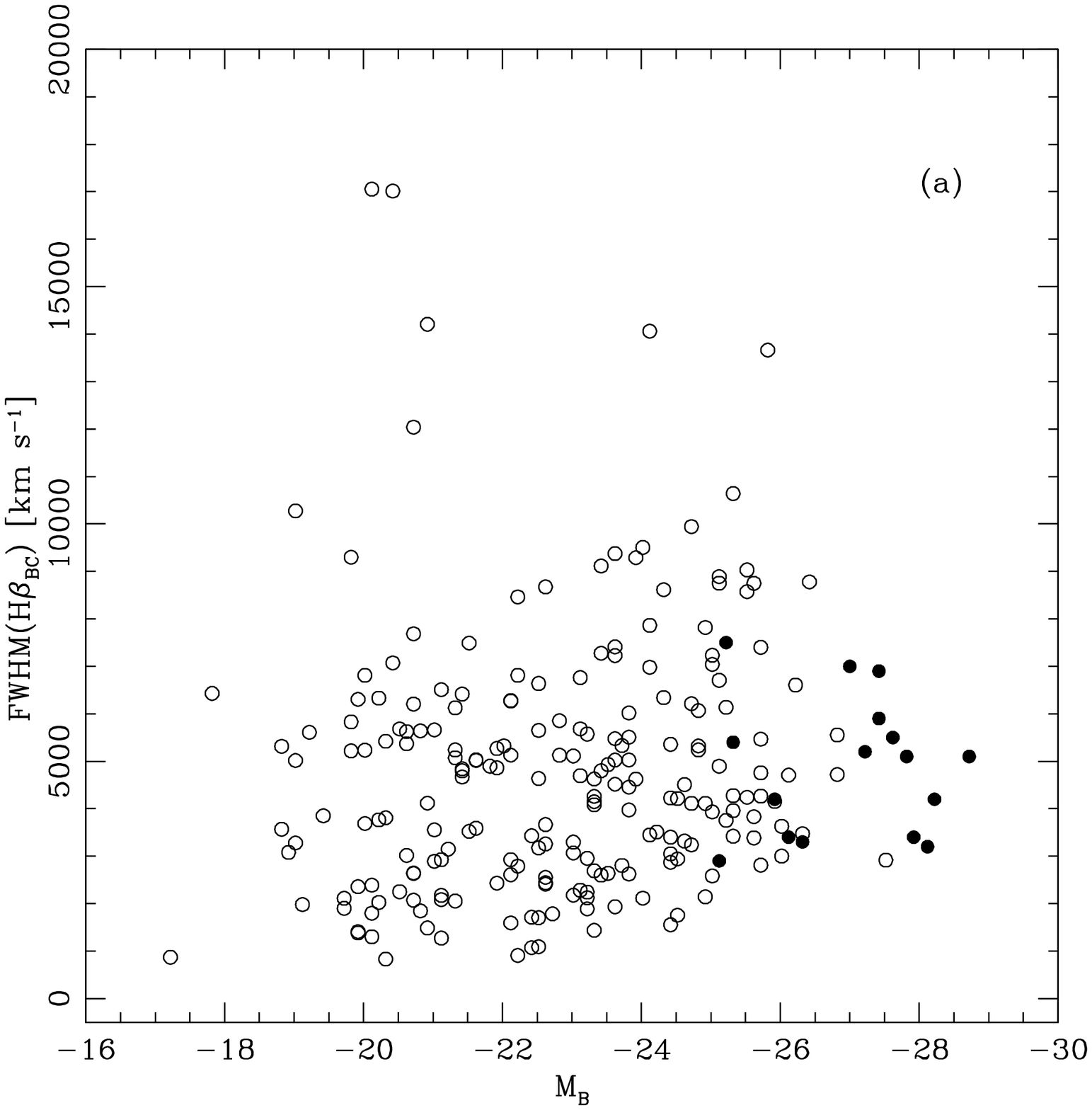}
  \includegraphics[width=8.0cm, height=8.0cm, angle=0]{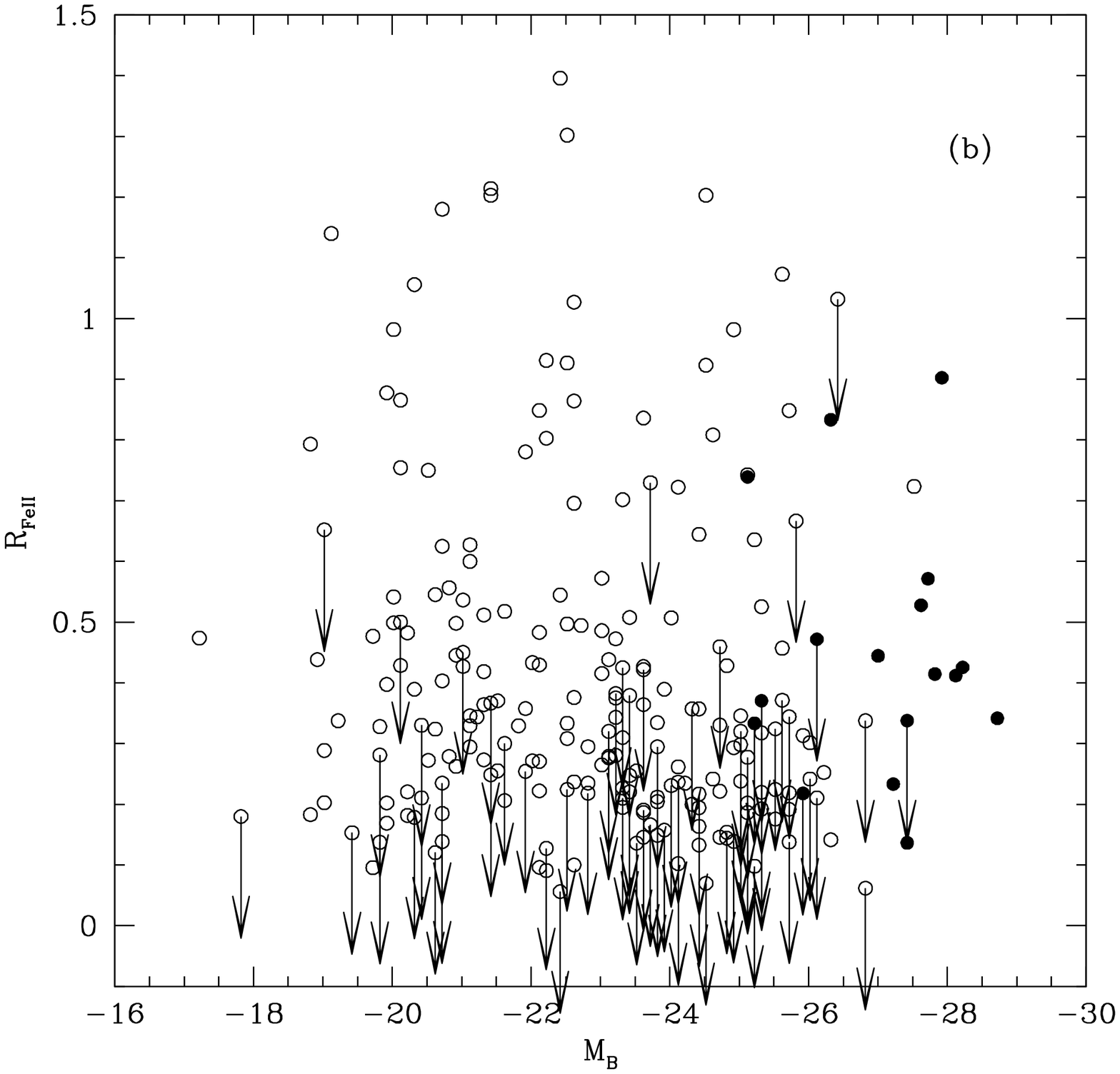}
  \includegraphics[width=8.0cm, height=8.0cm, angle=0]{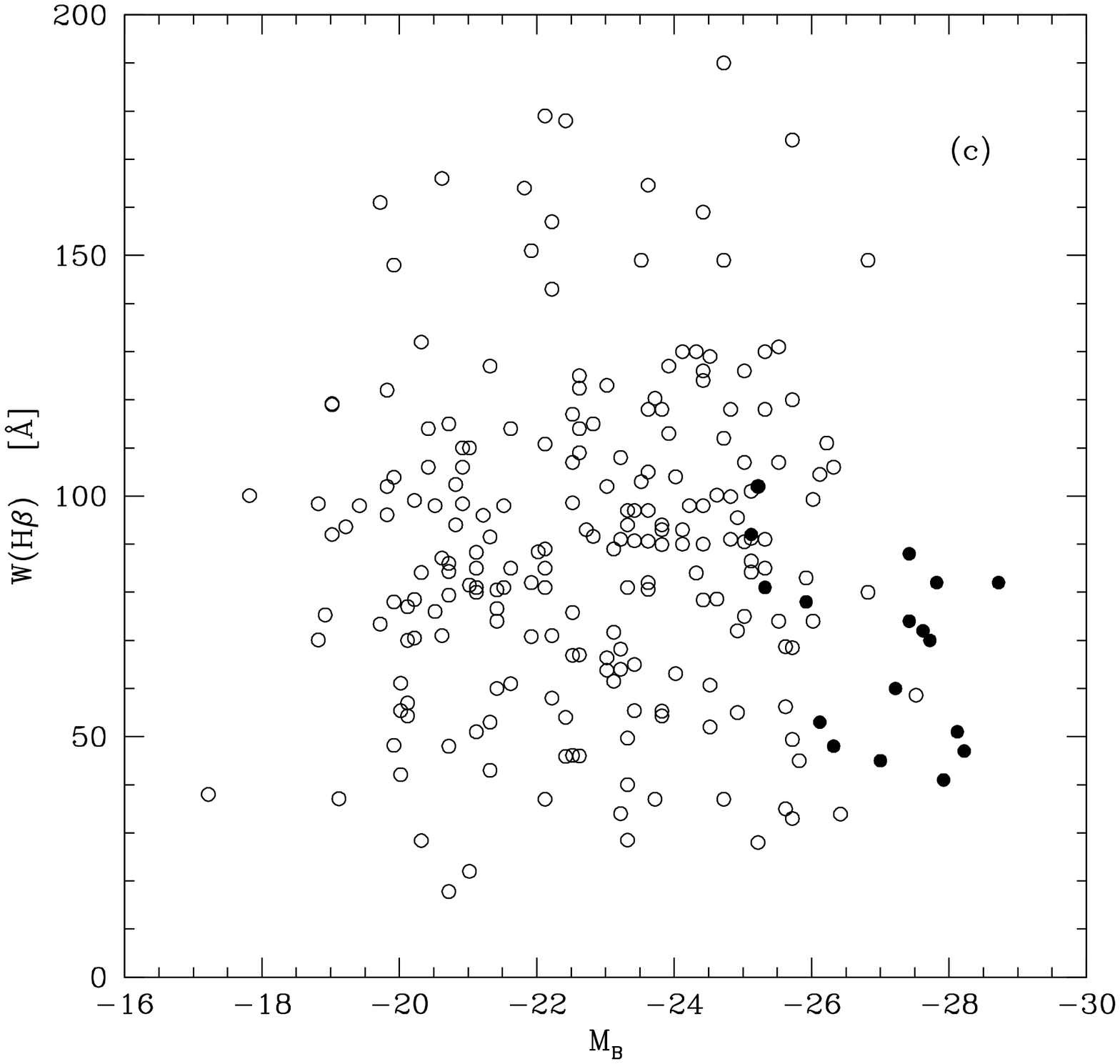}
  \includegraphics[width=8.0cm, height=8.0cm, angle=0]{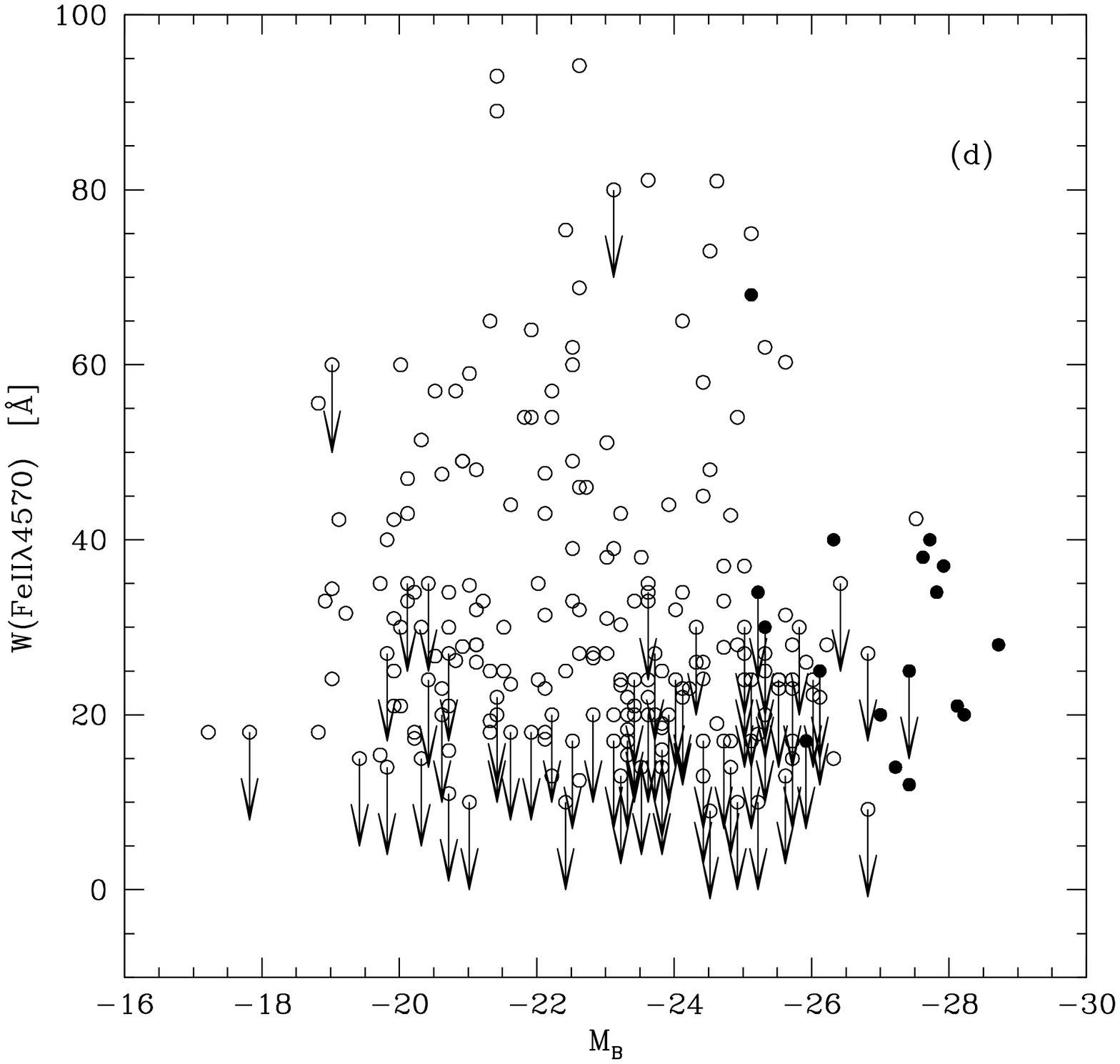}
 \includegraphics[width=8.0cm, height=8.0cm, angle=0]{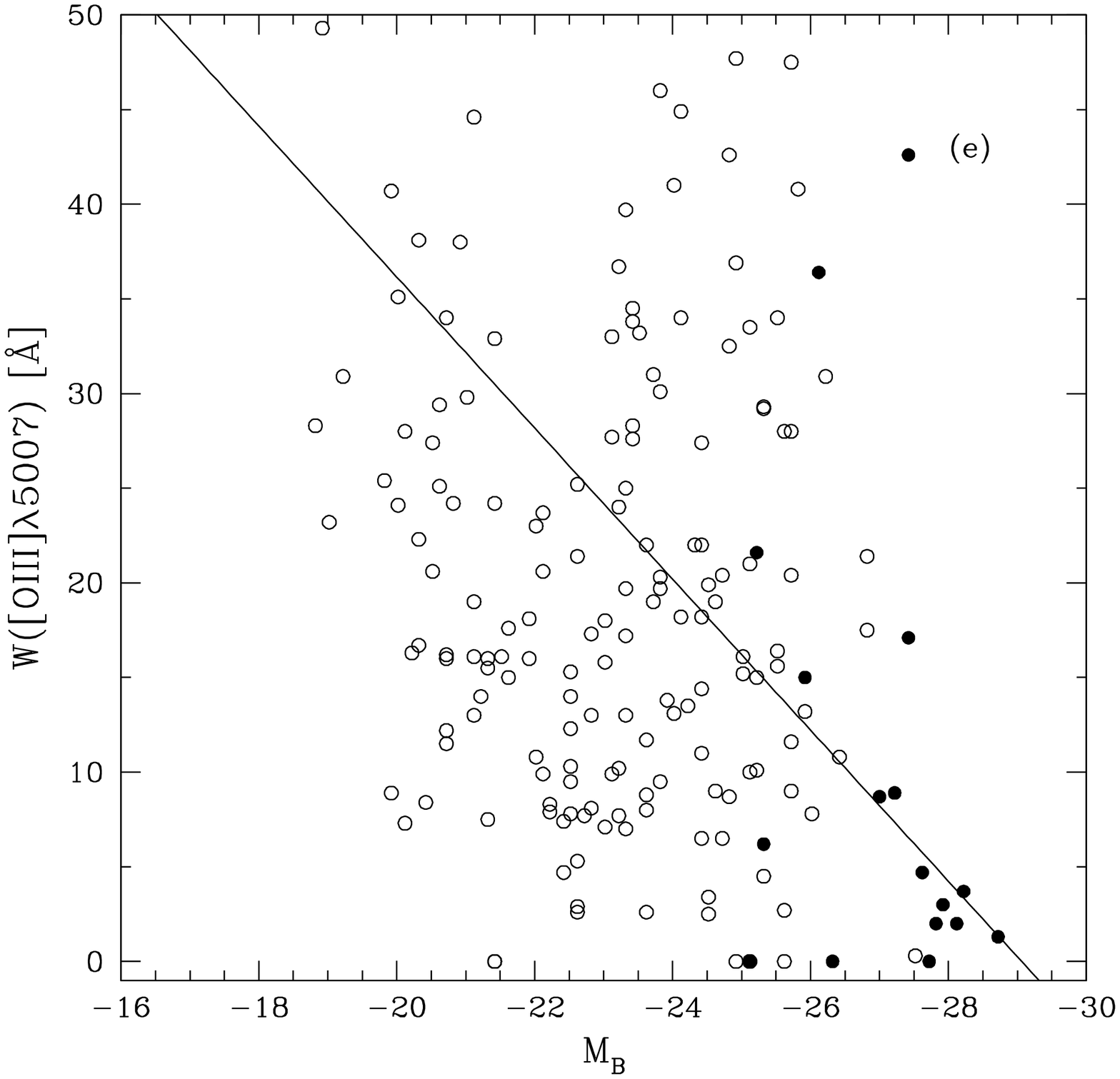}

      %\vspace{22cm}
      \caption[]{FWHM(\hbbc), \rfe, W(\hbbc), W(\feiiq), and W(\oiii) vs absolute blue magnitude \Mb. Equivalent widths are in the rest frame,
      in units of \AA; FWHM(\hbbc) is in \kms. Open circles: M03 data; filled circles: VLT/ISAAC sources. Best fit according to a robust method is provided in the only case of a significant correlation.}
         \label{fig:corr}
   \end{figure*}

\paragraph{FWHM(\hbbc):}  Figure \ref{fig:corr}a shows the M03b+VLT
source distributions in the FWHM(\hbbc) vs. \Mb\ plane. 
Early, as well as recent, works (e.g., Joly  et al.\ \cite{joly}; 
Corbett et al.\ \cite{corbett}) suggested at most a weak correlation  
between FWHM(\hb) and source luminosity. A weak tendency for 
FWHM(\hbbc) to increase with \Mb\ was also suspected in the M03b 
data. The weakness of the ``correlation'' is 
quantified by the Spearman correlation coefficients ($\approx-0.15$) 
for the M03b+VLT/ISAAC sample. Any  weak trend in this plot
disappears completely if we correct for a correlation 
induced because RL sources in our sample tend to have higher 
mean luminosity and FWHM(\hbbc) than RQ sources. 
The most significant feature of the plot involves an apparent 
systematic increase in the smallest observed FWHM with increasing \Mb.
FWHM(\hbbc) increases from 1000 to 3000 km~$^{-1}$ between \Mb~$= -23$
and $-28$. Very broad H$\beta$ profiles are observed at all
luminosities although sources with FWHM(\hbbc)~$>10^4$~km/s appear to
be quite rare and possibly disjoint with respect to the bulk of the
sample.  This sparseness is likely physically motivated because
sources with large FWHM(\hbbc) and W(\hbbc) really exist and are 
not the product of observational errors. Such sources are sometimes 
double-peaked (Eracleous \& Halpern \cite{eracleous}, and references therein)
and are also somewhat unstable systems  that may be radiating at 
very low L/M in the E1 context and may therefore be intrinsically 
short-lived (Sulentic et al.\ \cite{S95}; Marziani et al.\ \cite{M01}). 

\paragraph{\rfe:} The second E1 optical parameter \rfe\ shows 
no convincing evidence (Fig.~\ref{fig:corr}b) for a correlation if RQ
and RL sources are considered separately (RL are weaker \feii\
emitters than RQ sources, which accounts for the somewhat larger
correlation coefficient when no RQ-RL distinction is made). The \rfe\
situation is however more complicated because many W(\feiiq)
measurements are actually upper limits. We therefore considered two
cases: 1) we used the best fit \feiiq\ estimates from M03b + Table 2
and  2) we used minimum values of \feiiq\ detectability (which depend
on FWHM and S/N) to derive \feiiq\ upper limits. We then performed a
censored data analysis computing Kendall's $\tau$, in addition to
Spearman $r$, as an estimator of the correlation. Similar results are
found in the two cases ($\tau$ values are not reported).

\subsubsection{W(\hbbc) and the \hbbc\ Profile}

Claims that W(\hbbc) decreases with luminosity have been rather 
unconvincing due to small sample sizes and large intrinsic scatter 
(Sulentic et al.\ \cite{S00a}, and references therein (no); Mc99 (no);  
Wilkes et al.\ \cite{wilkes99} (yes); Croom et al.\ \cite{croom} (no)). 
In the M03b+VLT samples W(\hbbc) shows no significant 
correlation with \Mb\ even if RQ and  RL subsamples are combined (in contrast 
to \rfe\ and FWHM(\hbbc)). It is however intriguing that W(\hbbc) 
measured for VLT sources are all $\la 100$~\AA\ (Fig.~\ref{fig:corr}c).

Table \ref{tab:profs} reports line profile measures for 16/17
VLT sources. \object{HE 0353$-$3919} is excluded because \hbbc\ falls in
the gap between the sZ and Z bands allowing only a rough
estimate of FWHM(\hbbc)~$\sim 6000$~\kms. We considered the
luminosity dependence of AI, c(1/4),  and c(3/4) which are the 
most robust parameters (least affected by errors; M96). No 
significant correlation was found.

\subsubsection{\feii\ Emission Properties}

Figure \ref{fig:corr}d shows no evidence for a correlation
between \Mb\ and W(\feiiq). The absence of a W(\feiiq) trend 
is less significant because of the larger uncertainty associated with these
measures. In modeling and subtracting \feii\ emission we found 
no source that significantly deviated from the scaled and broadened 
\object{I~Zw~1} template. This template is remarkably successful for modeling 
even sources with strong and narrow \feii\ emission 
such as \object{HE 2305$-$5315} and \object{HE 0122$-$3759}. It also works well for sources
with obviously broader lines (e.g. \object{HE 0248$-$3628}). We found
no convincing examples of unusual \feii\ emission (i.e. multiplet 
ratios different from \object{I~Zw~1}).

\begin{figure*}
  \includegraphics[width=8.8cm, height=8.8cm, angle=0]{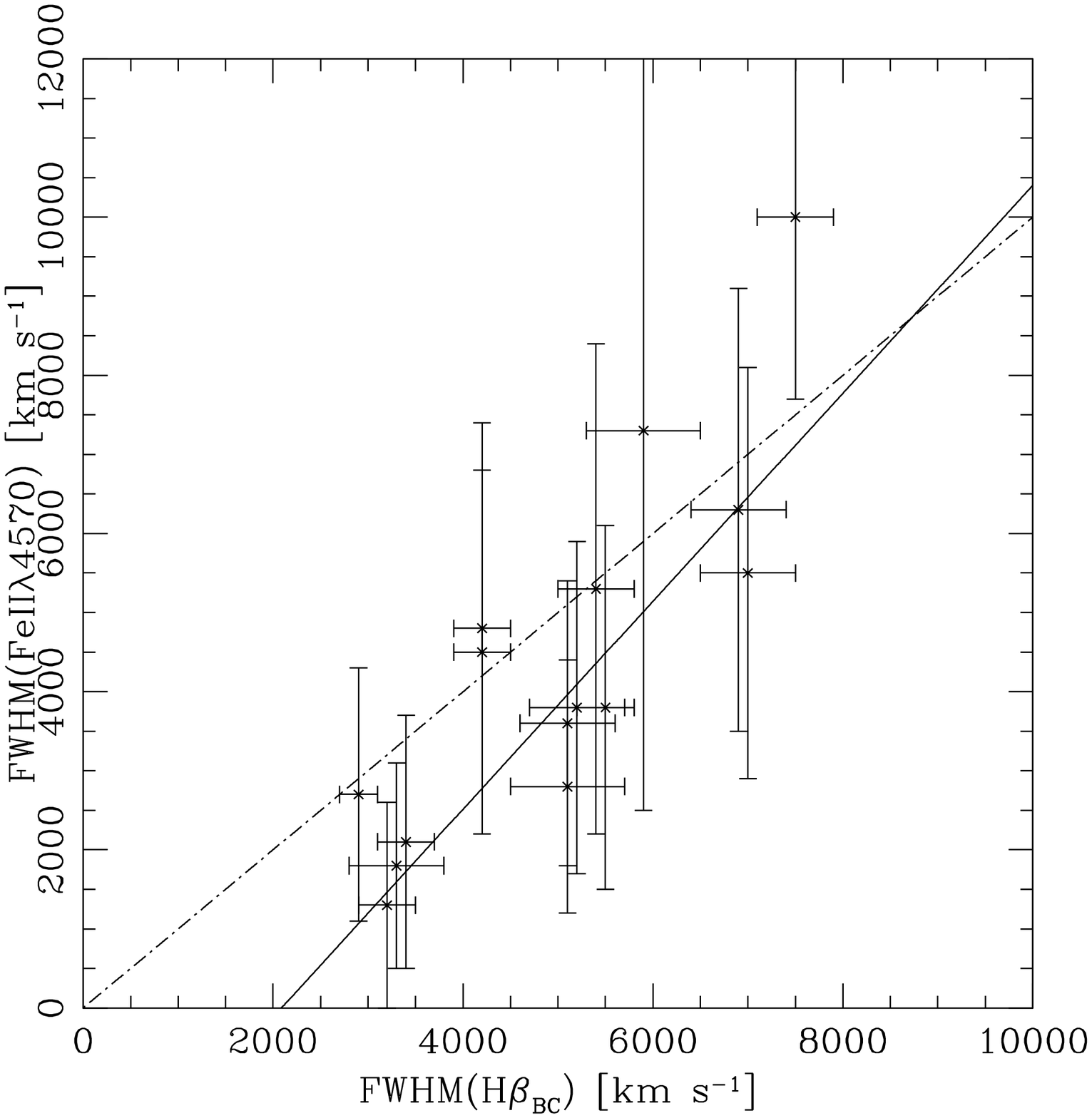}

      %\vspace{22cm}
      \caption[]{FWHM(\feiiq) vs.\ FWHM(\feiiq) for the 17 VLT/ISAAC sources.
      The dot-dashed line traces the locus of FWHM(\hbbc)=FWHM(\feiiq),
      while the continuous line is a least-squares best fit.}
         \label{fig:feii}
   \end{figure*}

FWHM(\hbbc) and FWHM(\feiiq) are strongly correlated as shown 
in Figure \ref{fig:feii}. This is consistent with the hypothesis 
that profiles of individual \feiiq\ lines are very similar to \hbbc. 
A weighted least-squares fit yields a slope of $\approx 1.31\pm0.22$
(1$\sigma$ uncertainty). The large uncertainty reflects
the relative insensitivity of the template fits to the adopted line width.

\subsubsection{Narrow Lines}
The VLT sample shows a large number of sources with weak \oiiiopt\
emission. Figure \ref{fig:corr}e shows
the best fit for all sources (M03+VLT/ISAAC) using a robust fitting 
method.  While this fit tells us little or nothing, we do see an 
interesting difference between the M03b and VLT samples. The M03b 
sample shows a large range in W(\oiiiopt) at all luminosities.
Most of the VLT measures cluster at very low values. The values are 
low even relative to the M03b sample that contains a significant number 
of NLSy1 sources. 
They are as low or lower than measured values for extreme 
Population A blue outlier sources (Marziani et al.\ \cite{M03a}). This may be
the first hint of a real decrease in the strength of the 
narrow line region for higher luminosity quasars. 

 \section{Discussion \label{disc}}

\subsection{The E1 Parameter Space \label{e1}}

The so-called E1 parameter space is a reasonable approximation to 
a (4D) H-R Diagram for AGN (S00b).
In this context we mean: 1) discrimination between and unification of
the diverse classes of AGN and, possibly, 2) representation of
various states of source spectral evolution. If an H-R analogy is in 
any way useful, it would not be surprising to find that more than two
observational parameters are required to define it. We take it as a
given that a quasar H-R Diagram is needed because of the striking diversity
in the spectral signatures of the broad line regions for different AGN
classes (S00b; Sulentic et al.\ \cite{S02}). It is hoped that the E1
parameter space will both clarify the phenomenology and better focus
models for the central geometry and physics. One of the big challenges
for E1 is to remove the degeneracy between physics and source
orientation to our line of sight (Sulentic et al.\ \cite{S03}). Right now, 
in analogy to mass (M) as the physical driver of the stellar main
sequence, evidence suggests  that the Eddington ratio ($\propto$~\.M)
is the principal physical driver in E1 (Marziani et al.\ \cite{M01}; M03b;
Boroson \cite{B02}). As an equivalent to the stellar main sequence we
find an ``L-shaped'' distribution of points in the
optical parameter plane of E1 (FWHM(\hbbc) vs.\ \rfe; Fig.
\ref{fig:e1}). The present tentative results suggest that 
the L-shaped distribution  is preserved up to \Mb$\approx-30$. 
We have earlier suggested that the extreme Population A sources with
the narrowest Balmer profiles, strongest FeII emission, CIV 
blueshift and soft X-ray excess are the high accretion end
of the E1 sequence (S00b). We also suggested that
these extreme objects may represent the quasar ``seed'' population (see also
Mathur \cite{mathur}). In this context we expect the fraction of such extreme
sources to increase with redshift. Both the SDSS CIV (Richards et al.\
\cite{richards}) and initial VLT samples are consistent with this idea (i.e.
high frequency of CIV blueshifts at high redshifts and weak narrow line regions
at intermediate redshifts, respectively). Overall however the VLT
sample follows closely the low-redshift E1 results.

\subsection{The Need for High S/N and Resolution\label{mc}}

The Mc99 data pose an apparent challenge to our claims 
about E1 robustness at higher redshifts. Unfortunately, the Mc99
data have very low S/N. Even if the spectral
resolution ($\approx530$~\kms) is modestly reasonable,
the S/N is in general $\la 10$ (see their \S 2 and Table 
\ref{tab:obs}). It is very risky to measure \feii\ emission 
in data with such low S/N and limited spectral coverage. 
M03b estimated  the minimum detectable W(\feiiq) as a function 
of FWHM(\hbbc) for several S/N values. For S/N~$\approx10$
we find the approximate relation: W(\feiiq)$_{\rm min} \approx
25 + 0.017\,$FWHM(\hbbc), where the rest frame W(\feiiq) is
in \AA, and FWHM(\hbbc) is expressed in \kms. W(\feiiq)$_{\rm 
min}\approx 50$\AA\ for FWHM(\hbbc)$\approx13000$~km~s$^{-1}$,
which explains why we have so many upper limits among the 
Population B sources. Most \rfe\ values reported by Mc99
should be changed to upper limits. We simulated Mc99 data with 
S/N~$\approx7$ (sources Q0049$+$007, Q0153$+$257, Q1011$+$091,
Q1309$-$056) with line widths and W(\feiiq) (always~$\ga60$~\AA)
as given in that paper. We then set the \rfe\ uncertainty to 
the upper limit W(\feiiq)$_{\rm min}$ needed for a
detection (i.e. upper limit is equal to the 3$\sigma$ uncertainty)
based on the errors in W(\hbbc) and FWHM(\hbbc) reported by Mc99. 
Even if \feiiopt\ is detected, the limited spectral coverage makes results
very sensitive to the somewhat arbitrary continuum placement (e.g.,
\object{Q1011$+$091}). Some results are obviously arbitrary (e.g. \object{Q0049$+$007} 
and \object{Q0226$-$104}) and there is no convincing evidence that
lines are very broad or that \feiiq\ should have W(\feiiq) $\sim$ 70
\AA. Another doubtful case involves \object{Q1209--056} although, again, the
limited spectral coverage makes \feii\ fitting an extrapolation for
the blue and red blends. If upper limits and revised uncertainties are
considered, the E1 quasar distribution in Fig.~\ref{fig:e1} can be 
significantly displaced toward \rfe~$\la0.5$. The same concerns apply to Yuan \&
Wills (\cite{yuan}), where both FWHM H$\beta$ and $R_{\rm FeII}$ values are
likely overestimated for many sources. Taken at face value
the Mc99 quasars would imply very large black hole masses (M$_{\rm BH}
\sim 10^{11}$~\msol) since they likely radiate at very low Eddington
ratio ($\sim0.01$--$0.1$) (Zamanov \& Marziani \cite{ZM02}). Such large
masses may not be frequent even among high-redshift quasars (McLure \&
Dunlop \cite{mclure}).

\subsection{A Luminosity/Mass Dependent Minimum FWHM(\hbbc)?\label{min}}

Two of our VLT Population A sources (\object{HE 0122$-$3759} and
\object{HE 2305$-$5315}) show \rfe\ (0.9 and 0.8) and no W(\oiiiopt) detections.
CIV data exists only for the former where an extreme CIV blueshift is
measured. We call such sources extreme Population A or NLSy1 sources
based on these criteria. The sources show FWHM(\hbbc) (3600 and
3200 \kms\ respectively)
which significantly exceeds  the nominal FWHM(\hbbc)~$\la 2000$~\kms\
limit for NLSy1 sources. Such broader-lined and strong \rfe\ Population
A sources are also found in our low-redshift sample. The smallest FWHM(\hbbc)
$\approx 2600$~\kms\ found among our 6 VLT
Population A sources is larger than the FWHM of more than half of the
Population A sources in our low-redshift sample. This suggests a
possible lower limit to this parameter that is rising with source
luminosity. HE 0122$-$3759 at $z\approx 2.4$, interpreted as an
NLSy1, would be the most luminous yet observed. No X-ray detections
are reported for these two sources.

\begin{figure*}
  \includegraphics[width=8.8cm, height=8.8cm, angle=0]{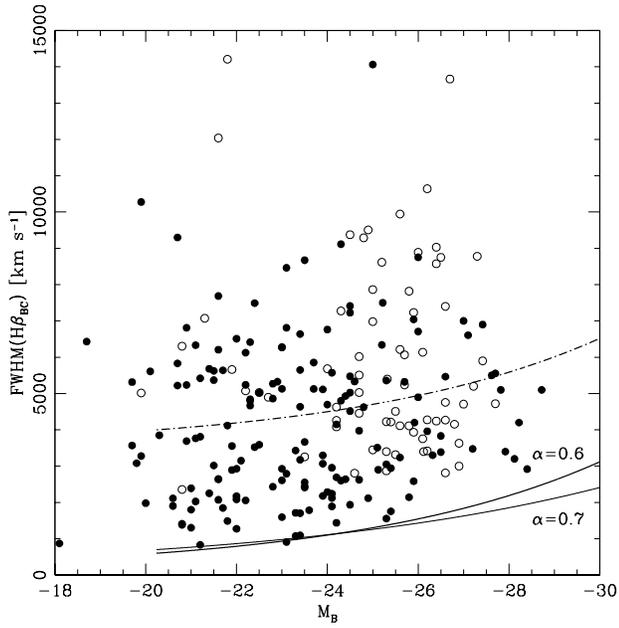}

      %\vspace{22cm}
      \caption[]{FWHM(\hbbc) vs. \Mb\ for the M03+VLT/ISAAC 232 sources, with RQ (filled circles) and RL (open circles)
      identified. The thick lines are the predicted minimum FWHM(\hbbc), as described in \S \ref{min} (solid line), and the
      boundary between Populations A and B (dot-dashed line). The thin line refers to a slightly different fit to the $R_{\rm BLR}-L$ relationship,
      with $\alpha=0.7$.}
         \label{fig:rqrl}
   \end{figure*}

In order to ascertain whether there is a luminosity effect, one
can consider the well-defined lower boundary in the FWHM(\hbbc) vs.\
\Mb\ diagram (Figs.~\ref{fig:corr}a and \ref{fig:rqrl}). This boundary
can be interpreted as a luminosity effect. Such a trend is indeed
expected if: 1) \hbbc\ broadening is dominated by virial motions and 2)
the emissivity-weighted distance of the BLR from the central BH
depends on luminosity $R_{\rm BLR} \propto L^{-\alpha}$\ (Kaspi et
al.\ \cite{kaspi}). The exact value of $\alpha$\ is very sensitive to: a)
the poor sampling in some luminosity ranges, b) the presence of several
outliers and c) the cosmology (Marziani et al.\ \cite{M03a}). Refitting
Kaspi's data for $H_0 = 75$ \kms~Mpc$^{-1}$\ and $q_0 = 0$, we
obtain:
$$ R_{\rm BLR} \approx 1.19 10^{17} (\frac{L}{10^{45}})^{0.60} {\rm cm}$$
where we have assumed that the bolometric luminosity $L \approx 10 \lambda
L_\lambda$\ at 5100 \AA. The value of $\alpha=0.7$  is slightly
different from the value given by Kaspi et al.\ (\cite{kaspi}).

The virial relationship implies:
$$
\frac{\sqrt{3}}{2} {\rm FWHM} = G^\frac{1}{2} M^\frac{1}{2} R_{\rm
BLR}^{-\frac{1}{2}}
$$
where the factor 1/2 comes from the use of FWHM(\hbbc) ($\Delta
v=\frac{1}{2}$FWHM) and the factor
$\sqrt{3}$ takes into account that we measure a radially
projected velocity. Substituting $R_{\rm BLR}$ with $L$ and
transforming into convenient units, we obtain:
$${\rm FWHM(H\beta_{\rm BC})} \approx 2880 L_{\odot,11}^{0.20}
(\frac{L}{M})^{-\frac{1}{2}}_{\odot,3} {\rm ~~km ~s^{-1}}$$
where luminosity is in $10^{11}$ L$_\odot$ and the
L/M ratio is in units of $10^3$ times the solar value
(L/L$_{\rm Edd}=1$  corresponds to $\log (\frac{L}{M})_\odot
\approx 4.53$).

We add here the assumption that low-redshift NLSy1
with the narrowest lines radiate very close to the Eddington ratio.
If we assume $\log \frac{L}{M} \approx4.5$ we obtain FWHM(\hbbc)$_{\rm
min} \approx600$~\kms\ for $\log L = 11$. The same relationship
written for $\log L = 11$ as a function of \Mb\ becomes:
\begin{eqnarray*}
{\rm FWHM_{\rm min}(H\beta_{\rm BC})} & \approx & 500 \times
10^{[-0.08(M_{\rm B} + 20.24)]}+100~{\rm km ~s^{-1}} \\
& \propto & 10^{(-0.08 M_{\rm B})}
\end{eqnarray*}
Considering the typical instrumental width of our data the actual
FWHM(\hbbc)$_{\rm min}$ would be $\sim
1000$~\kms. FWHM(\hbbc)$_{\rm min}$\ is similar to the lowest
FWHM(\hbbc) found in the M03b sample. It is even closer to the
lowest known FWHM(\hbbc) for NLSy1s (\object{PHL~1092} and \object{IRAS
13224$-$3809}; Marziani et al.\ \cite{M01}). If we consider the luminosity
dependence of FWHM(\hbbc)$_{\rm min}$, we see that the expected trend
for $\alpha = 0.6$\ reproduces fairly well the FWHM(\hbbc) lower
boundary as a function of \Mb, especially if we consider only RQ
sources (see Fig. \ref{fig:rqrl}). A less pronounced trend is 
expected for $\alpha=0.7$, especially at  high luminosity. This result
is helpful for interpreting the following three issues:

\paragraph{Correlation of FWHM(\hbbc) with luminosity:}
A significant correlation between FWHM and luminosity 
may depend on: 1) sample selection and 2) intrinsic dispersion of 
FWHM values in a narrow \Mb\ range. It will also be affected by the 
fact that the profile of H$\beta$ is now known to be composed of 
at least three independent components; narrow, broad and very broad
(Sulentic et al.\ \cite{S02}). Population A sources appear to be dominated 
by the ``classical'' broad component. Population B can be significantly 
affected by the unshifted narrow and redshifted very broad components.
In many sources these two components can cancel any bias on the 
measured FWHM \hbbc. In others they do not and resultant FWHM measures
cannot be safely  compared with the above or Population A sources.
Given the low S/N of most
quasar spectra, the very broad component will often be modelled with 
the continuum. FWHM measures for such sources will often be serious 
underestimates unless the narrow component is explicitly subtracted.  
This leads to a prediction that RL sources (mostly Population B) will
be systematically measured with FWHM (and consequently, M$_{BH}$) too 
low in low S/N spectra. 

One must consider that any expected FWHM-luminosity dependence 
will be  very weak. This means that it is reasonable to expect an increase
$\Delta$FWHM(\hbbc$)\approx 1000$~\kms\  over an increase of
$\Delta$\Mb~$\approx 10$, with FWHM(\hbbc)$_{\rm min}$ changing from
1000~\kms\ to 2000~\kms. In a narrow \Mb\ range, the intrinsic
spread of FWHM(\hbbc) measures ranges from 1000--10000~\kms.
This will tend to make any intrinsic correlation very weak.
Larger samples at higher luminosity are needed to test these
predictions. Corbett et al.\ (\cite{corbett}), analyzing a very large sample
from the 2dF and 6dF redshift surveys, find a weak increase of \hb\
line width with luminosity, with a slope $\approx1.5$--0.2 ($\Delta v$
vs.\ $\log L$), very close to the one expected from our calculation.
However the average resolution, S/N and lack of FeII subtraction
in this analysis warrant caution in accepting this result as support 
for our prediction. Higher luminosity/redshift spectra in this sample 
will show systematically lower S/N. If FeII-strong Population A
sources really dominate at high redshift then most sources will have
FWHM overestimated due in part to FeII blending with H$\beta$.
Comparable (and reasonably high) S/N data are needed for sources at
all redshifts and luminosities in order to make a proper luminosity
correlation test.

\paragraph{RQ and RL differences:} The FWHM(\hbbc) difference between 
RQ and RL sources is not simply a luminosity effect: if redshift and 
magnitude distributions are matched, RQ and RL still show systematic differences
(M03b; Sulentic et al.\ \cite{S03}). This can also be seen
restricting attention to higher luminosities (\Mb~$\la-24$)
where most RL are located: in this range FWHM(\hbbc)$_{\rm RL}
\ga $~FWHM(\hbbc)$_{\rm RQ}$\ (a Kolmogorov-Smirnov test indicates
probability $\rm P\sim0.01$ that the FWHM values are drawn from the
same distribution).

\paragraph{Boundaries: Populations A-B and NLSy1s:}
If we accept a low-redshift boundary between Population A and B
sources at FWHM(\hbbc)~$\approx4000$~\kms, then it may
increase as a function of luminosity following a curve parallel to that for
FWHM(\hbbc)$_{\rm min}$. If this correction is applied, the frequency
of Population A (10) and Population B (7) sources among the 17
VLT/ISAAC sources is consistent with the low-redshift sample
(Population A should be $\sim60$--$65$\% of all sources; Marziani et
al.\ \cite{M03a}).

\section{Summary and Conclusion\label{concl}}

We present 17 VLT/ISAAC spectra of intermediate redshift quasars
with resolution and S/N comparable to our M03b sample of
ground-based spectra for low-redshift AGN. Quasar spectra in
the $1\la z \la 2.5$ range do not yet show appreciable E1 parameter
differences  from quasars with $z<1.0$. No significant luminosity
correlations with line parameters are found again in agreement
with previous E1 inferences. The two most interesting  effects found in
this preliminary comparison involve: 1) a tendency for the minimum FWHM
H$\beta$ to increase from $\sim1000$~km~s$^{-1}$ at \Mb~$= -20$ to
$\sim3000$~km~s$^{-1}$ at \Mb~$= -28$ and 2) most of our VLT sources
show W(\oiiiopt) values equal to or less than those found for the 
low-redshift sample.  The former effect can be accounted for by
accretion theory while the latter may indicate a weakening of the narrow 
line region in higher redshift quasars.  A comparison with the largest 
sample of previously published IR data (Mc99) indicates that high 
S/N and spectral resolution are required to obtain meaningful results. 
Further VLT/ISAAC observations will provide a unique window on the E1
parameter space at high luminosity and redshift.

\begin{acknowledgements}
We thank Lutz Wisotzki for providing us with the HE optical spectra.

\end{acknowledgements}

\begin{table*}
\begin{center}
\caption{Basic Properties of Sources and Log of Observations \label{tab:obs}}
    \begin{tabular}{lllllllllllll}
    \hline  \hline
    \noalign{\smallskip}
     Object name & \multicolumn{1}{c}{$\rm m_B^{\mathrm{a}}$}
     &\multicolumn{1}{c}{$z^{\mathrm{b}}$} &
     \multicolumn{1}{c}{Line$^{\mathrm{c}}$} &\multicolumn{1}{c}{$M_{\mathrm B}^
     {\mathrm{d}}$}  & $\log \rm R_K^{\mathrm{e}}$ & Date$^{\mathrm{f}}$
     &
\multicolumn{1}{c}{Band$^{\mathrm{g}}$} &
\multicolumn{1}{c}{DIT$^{\mathrm{h}}$}
     & \multicolumn{1}{c}{N$_{\rm exp}^{\mathrm{i}}$} & \multicolumn{1}{c}{Airmass$^{\mathrm{j}}$} &
     \multicolumn{1}{c}{S/N$^{\mathrm{k}}$} \\
%     \multicolumn{1}{c}{} & \multicolumn{1}{c}{} & \multicolumn{1}{c}{\AA} & \multicolumn{1}{c}{\AA}
%     &
\multicolumn{1}{c}{(1)}      & \multicolumn{1}{c}{(2)}         &
\multicolumn{1}{c}{(3)}
     & \multicolumn{1}{c}{(4)}    & \multicolumn{1}{c}{(5)} & \multicolumn{1}{c}{(6)}  & \multicolumn{1}{c}{(7)}  & \multicolumn{1}{c}{(8)} & \multicolumn{1}{c}{(9)} & \multicolumn{1}{c}{(10)} & \multicolumn{1}{c}{(11)} & \multicolumn{1}{c}{(12)}\\
     \hline
     \noalign{\smallskip}
\object{HE0003$-$5026}&  17.07 &1.0772(4)  &1  & $-$26.7   &2.19 & 2001-12-23 & Z    & 180   & 12  & 1.42-1.59 & 20\\
             &        &           &   &         &    &            & sZ   & 180   & 12  & 1.27-1.38 & $\sim 5$ \\
\object{HE0005$-$2355}&  16.94 & 1.4120(3) &2  & $-$27.6   &2.56 &2001-11-22  & J    & 120   & 24  & 1.01-1.07 &15 \\
\object{HE0048$-$2804}&  17.25 & 0.8467(3)&1  & $-$26.0   & \ldots     &2001-12-17  & Z    & 150   & 16  & 1.04-1.11 & 7\\
\object{HE0122$-$3759}&  16.94 & 2.2004(4) &2  & $-$28.9   &  \ldots   &2001-12-28  & SH   & 180   & 12  & 1.05-1.09 & 15\\
\object{HE0205$-$3756}&  17.17 & 2.4367(5) &2  & $-$29.0   & \ldots    &2002-12-16  & SH   & 180   & 12  & 1.05-1.09 & 35\\
\object{HE0248$-$3628}&  16.58 & 1.5362(4) &2  & $-$28.2   &0.84 &2001-12-28  & J    & 180   & 12  & 1.03-1.05 & 30\\
\object{HE0331$-$4112}&  16.24 & 1.1153(4) &1,2& $-$27.6   & \ldots    &2001-12-23  & Z    & 180   & 12  & 1.25-1.39 & 30\\
             &        &           &   &         &      &            &sZ    & 180   & 8   & 1.41-1.54 & 15\\
\object{HE0349$-$5249}&  16.13 & 1.5384(4) &2  & $-$28.7   & \ldots    &2002-02-26  & J    & 120   & 20  & 1.26-1.38 & 30\\
\object{HE0353$-$3919}&  16.14 & 1.0065(35)&3  & $-$27.5   &1.49 &2002-02-26  & Z    & 180   & 12  & 1.41-1.62 & 30\\
             &        &           &   &         &      &            &sZ    & 180   & 12  & 1.70-2.05 & 15\\
\object{HE0454$-$4620}&  17.23 & 0.8528(3) &1  & $-$25.9   &3.37 &2002-12-16  & Z    & 150   & 18  & 1.25-1.40 &  5\\
\object{HE2202$-$2557}&  16.71 & 1.5347(3) &2  & $-$28.1   &1.80 &2001-10-05  & J    & 120   & 12  & 1.06-1.03 & 20\\
\object{HE2259$-$5524}&  17.09 & 0.8549(4) &2  & $-$26.1   & \ldots    &2001-10-05  & Z    & 180   & 18  & 1.20-1.16 &  10\\
\object{HE2305$-$5315}&  16.33 & 1.0733(4) &2  & $-$27.5   & \ldots    &2001-11-24  & Z    & 120   & 12  & 1.21-1.25 & 35 \\
             &        &           &   &         &      &            &sZ    & 180   & 8   & 1.17-1.20 &  10\\
\object{HE2340$-$4443}&  17.07 & 0.9216(3) &1,2&  $-$26.3  & \ldots    &2001-11-25  & Z    & 180   & 20  & 1.07-1.13 & 20--5 \\
\object{HE2349$-$3800}&  17.46 & 1.6040(4) &2,4&  $-$27.4  &1.93 &2001-11-25  & J    & 180   & 12  & 1.10-1.17 & 35--15 \\
\object{HE2352$-$4010}&  16.05 & 1.5799(4) &2  &  $-$28.8  & \ldots    &2001-10-05  & J    & 180   & 12  & 1.04-1.03 & 60--35 \\
\object{HE2355$-$4621}&  17.13 & 2.3825(3) &1,2&  $-$28.9  & \ldots    &2001-11-24  & SH   & 180   & 24  & 1.13-1.29 & 20 \\
\noalign{\smallskip} \hline \hline
\end{tabular}
\end{center}
\begin{list}{}{}
\item[$^{\mathrm{a}}$] Apparent B magnitude.
\item[$^{\mathrm{b}}$] Redshift, with uncertainty in parenthesis.
\item[$^{\mathrm{c}}$] Lines used for redshift calculations: 1: \oiii, 2: \hb, 3: \feiiq, 4: H$\gamma$.
\item[$^{\mathrm{d}}$] Absolute B magnitude, computed for $H_0$=75 \kms Mpc$^{-1}$,  $q_0$=0, and $k$-correction
spectral index $\alpha$=0.6. Note that  \Mb\ values are not those
reported in the Ver\'on-Cetty \& V\'eron (\cite{veron}) catalogue, but
have been computed from the apparent B magnitude provided in the HE
survey tables.
\item[$^{\mathrm{e}}$] Decimal logarithm of the
specific flux ratio at 6cm and 4400 \AA\ (effective wavelength of  the
B band)
\item[$^{\mathrm{f}}$] Date refers  to time at start of
exposure.
\item[$^{\mathrm{g}}$] Photometric band.
\item[$^{\mathrm{h}}$] Detector Integration Time (DIT) of ISAAC, in
seconds.
\item[$^{\mathrm{i}}$] Number of exposures with single
exposure time equal to DIT.
\item[$^{\mathrm{j}}$] Airmass at  start and end of exposure.
\item[$^{\mathrm{k}}$] S/N at continuum level. Where
two values are reported they are for the blue and red side of \hbbc\
respectively. The S/N value  is with N estimated  at a $2 \sigma$\
confidence level i.e., 2rms.
\end{list}
\end{table*}

\begin{table*}
\begin{center}
\caption{Measurements of Equivalent Widths and FWHM of Strongest Lines} \label{tab:meas}
    \begin{tabular}{llllllll}
    \hline  \hline
    \noalign{\smallskip}
     Object name& \multicolumn{1}{c}{W(\hbbc)$^{\mathrm{a}}$} &
  \multicolumn{1}{c}{W(\feiiq)$^{\mathrm{b}}$}
     & \multicolumn{1}{c}{FWHM(\feiiq)$^{\mathrm{c}}$}
      &\multicolumn{1}{c}{W(\hbnc)$^{\mathrm{d}}$}
     & \multicolumn{1}{c}{W({\sc[Oiii]$\lambda4959$})$^{\mathrm{d}}$}
     & \multicolumn{1}{c}{W(\oiii)$^{\mathrm{d}}$}
    \\
%     \multicolumn{1}{c}{} & \multicolumn{1}{c}{} & \multicolumn{1}{c}{\AA} & \multicolumn{1}{c}{\AA}
%     &
\multicolumn{1}{c}{(1)}      & \multicolumn{1}{c}{(2)}         &
\multicolumn{1}{c}{(3)}
     & \multicolumn{1}{c}{(4)}    & \multicolumn{1}{c}{(5)} & \multicolumn{1}{c}{(6)} & \multicolumn{1}{c}{(7)}
       \\
     \hline
     \noalign{\smallskip}
\object{HE0003$-$5026} & 81 $\pm$9   &  25 $\pm$10&5300  $\pm$  3100 &0.5$\pm$0.3  & 0.98$\pm$0.4  & 6.2 $\pm$0.7 &   \\
\object{HE0005$-$2355} & 74 $\pm$7   &  11 $\pm$7 &7300  $\pm$  4800 &4.0$\pm$0.6  & 11.9$\pm$1.2  & 42.6$\pm$4  &   \\
\object{HE0048$-$2804} & 102$\pm$10 &   40 $\pm$15&10000 $\pm$  2300 &1.8$\pm$0.9  & 6.5$\pm$1     & 21.6$\pm$2.2&   \\
\object{HE0122$-$3759} & 41 $\pm$4   &  37 $\pm$5 &2100  $\pm$  1600 & 0           & 0$\pm$1       & 3.0 $\pm$1   &   \\
\object{HE0205$-$3756} & 82 $\pm$10  &  34 $\pm$5 &3600  $\pm$  1800 &0.8$\pm$0.8  & 0.4$\pm$0.3   & 2.0 $\pm$1.5 &   \\
\object{HE0248$-$3628} & 47 $\pm$5   &  20 $\pm$3 &4500  $\pm$  2300 &1.1$\pm$0.4  & \ldots$^{\mathrm{e}}$          & 3.7 $\pm$1   &   \\
\object{HE0331$-$4112} & 72 $\pm$8   &  38 $\pm$7 &3800  $\pm$  2300 &0.5$\pm$0.5  & 2.1$\pm$0.5   & 4.7 $\pm$0.5 &   \\
\object{HE0349$-$5249} & 82 $\pm$9   &  28 $\pm$3 &2800  $\pm$  1600 &0.4$\pm$0.4  & 0$\pm$1       & 1.3 $\pm$0.6 &   \\
\object{HE0353$-$3919} & 70 $\pm$20  &  40 $\pm$20&4800  $\pm$  2600 & 0           & 0             & 0           &   \\
\object{HE0454$-$4620} & 53 $\pm$6   &  10 $\pm$10&   \ldots          & 0           & 13.7$\pm$2   & 36.4$\pm$4  &   \\
\object{HE2202$-$2557} & 45 $\pm$5   &  20 $\pm$3 &5500 $\pm$   2600 &1.25$\pm$0.4 & 3$\pm$1       & 8.7 $\pm$0.9&   \\
\object{HE2259$-$5524} & 92 $\pm$10  &  68 $\pm$15&2700 $\pm$   1600 & 0           & 0             & 0           &   \\
\object{HE2305$-$5315} & 48 $\pm$5   &  40 $\pm$15&1800 $\pm$   1300 & 0           & 0             & 0           &   \\
\object{HE2340$-$4443} & 78 $\pm$8   &  13 $\pm$3 &4800 $\pm$   2600 &2.4$\pm$0.8  & 6.6$\pm$  0.7 & 15  $\pm$1.5&   \\
\object{HE2349$-$3800} & 60 $\pm$6   &  24 $\pm$4 &3800 $\pm$   2100 &2.1$\pm$0.7  & 1.0$\pm$  0.2 & 8.9 $\pm$0.9 &   \\
\object{HE2352$-$4010} & 51 $\pm$5   &  21 $\pm$3 &1300 $\pm$   1300 & 0           & 0             & 2.0 $\pm$1.0 &   \\
\object{HE2355$-$4621} & 88 $\pm$9   &  12 $\pm$2 &6300 $\pm$   2800 &2.4$\pm$0.8  & 3.75$\pm$0.5  & 17.1$\pm$2  &   \\
\noalign{\smallskip} \hline \hline
\end{tabular}
\end{center}
\begin{list}{}{}
\item[$^{\mathrm{a}}$] Rest frame equivalent width of \hbbc\ in \AA\ $\pm 2\sigma$ confidence level uncertainty.
\item[$^{\mathrm{b}}$] Rest frame equivalent width of the \feiiq\ blend in \AA\ $\pm 2\sigma$ confidence level uncertainty.
\item[$^{\mathrm{c}}$] FWHM of lines in the \feiiq\ blend and uncertainty at 2$\sigma$, in \kms. See text for details.
\item[$^{\mathrm{d}}$] Rest frame equivalent width of \hbnc, {\sc [Oiii]}\l 4959, and \oiii\ in \AA, with
uncertainty at 2$\sigma$.
\item[$^{\mathrm{e}}$] {\sc [Oiii]}$\lambda$4959 not visible. \oiii\ is instead well visible.
\end{list}
\end{table*}

\begin{table*}
\begin{center}
\caption{\hbbc\ Line Profile Measurements} \label{tab:profs}
    \begin{tabular}{lllllllllllllllll}
    \hline  \hline
    \noalign{\smallskip}
     \multicolumn{1}{c}{Source}
     & \multicolumn{1}{c}{FWZI$^{\mathrm{a}}$}
     & \multicolumn{1}{c}{$\Delta^{\mathrm{a,b}}$}
     & \multicolumn{1}{c}{FWHM$^{\mathrm{a}}$}
     & \multicolumn{1}{c}{$\Delta^{\mathrm{a,b}}$}
     & \multicolumn{1}{c}{A.I.$^{\mathrm{c}}$}
     & \multicolumn{1}{c}{$\Delta^{\mathrm{b}}$}
     & \multicolumn{1}{c}{Kurt.$^{\mathrm{d}}$}
     & \multicolumn{1}{c}{$\Delta^{\mathrm{b}}$}
     & \multicolumn{1}{c}{c(1/4)}
     & \multicolumn{1}{c}{$\Delta^{\mathrm{a,b}}$}
     & \multicolumn{1}{c}{c(1/2)$^{\mathrm{a}}$}
     & \multicolumn{1}{c}{$\Delta^{\mathrm{a,b}}$}
     & \multicolumn{1}{c}{c(3/4)$^{\mathrm{a}}$}
     & \multicolumn{1}{c}{$\Delta^{\mathrm{a,b}}$}
     & \multicolumn{1}{c}{c(0.9)$^{\mathrm{a}}$}
     & \multicolumn{1}{c}{$\Delta^{\mathrm{a,b}}$}
     \\
\multicolumn{1}{c}{(1)}      & \multicolumn{1}{c}{(2)}         &
\multicolumn{1}{c}{(3)}
     & \multicolumn{1}{c}{(4)}    & \multicolumn{1}{c}{(5)} & \multicolumn{1}{c}{(6)} & \multicolumn{1}{c}{(7)}
     & \multicolumn{1}{c}{(8)}   & \multicolumn{1}{c}{(9)}  & \multicolumn{1}{c}{(10)}
      & \multicolumn{1}{c}{(11)}  & \multicolumn{1}{c}{(12)}  & \multicolumn{1}{c}{(13)}
       & \multicolumn{1}{c}{(14)}  & \multicolumn{1}{c}{(15)}  & \multicolumn{1}{c}{(16)}
        & \multicolumn{1}{c}{(17)}   \\    \hline
     \noalign{\smallskip}
\object{HE0003$-$5026} & 20500& 400  & 5400  & 400 &   $-$0.02 & 0.10  & 0.32 &   0.05 &  $-$1300  &900 &$-$250   & 410 &$-$520  &  270& $-$520  &  180\\
\object{HE0005$-$2355} & 20100& 3100  &5900  & 600 &   0.27  & 0.06  & 0.28 &   0.03 &  3000   &800 &1180   & 620 &410   &  320& 20    &  190\\
\object{HE0048$-$2804} & 19700& 3400  &7500  & 400 &   0.04  & 0.10  & 0.44 &   0.06 &  300    &1100&590    & 390 &820   &  420& $-$90   &  360 \\
\object{HE0122$-$3759} & 12100& 2800  &3400  & 300 &   $-$0.08 & 0.06  & 0.33 &   0.04 &  $-$500   &300 &$-$100   & 300 &$-$30   &  170& $-$20   &  110\\
\object{HE0205$-$3756} & 22200& 1800  &5100  & 500 &   0.34  & 0.08  & 0.22 &   0.04 &  2600   &1200&$-$310   & 480 &$-$680  &  170& $-$510  &  130\\
\object{HE0248$-$3628} & 21200& 1200  &4200  & 300 &   $-$0.12 & 0.16  & 0.30 &   0.08 &  $-$800   &1200&$-$330   & 340 &$-$20   &  230& 30    &  140\\
\object{HE0331$-$4112} & 17700& 1300  &5500  & 300 &   0.00  & 0.11  & 0.39 &   0.07 &  1600   &1000&990    & 310 &840   &  280& 820   &  210\\
\object{HE0349$-$5249} & 24700& 2300  &5100  & 600 &   0.46  & 0.08  & 0.20 &   0.04 &  7800   &1900&1110   & 610 &670   &  260& 520   &  170\\
\object{HE0454$-$4620} &  7700&  300  &3400  & 200 &   $-$0.20 & 0.08  & 0.43 &   0.05 &  1500  & 300        & 1120&250  &\ldots&  150& 160&  110\\
\object{HE2202$-$2557} & 22600& 1700  &7000  & 500 &   0.16  & 0.10  & 0.34 &   0.06 &  1800   &1300&930    & 490 &280   &  360& $-$80   &  270\\
\object{HE2259$-$5524} & 9700&  1000  &2900  & 200 &   $-$0.17 & 0.10  & 0.31 &   0.06 &  $-$800   &500 &$-$90    & 180 &70    &  130& 20    &   80\\
\object{HE2305$-$5315} & 15000& 4900  &3300  & 500 &   $-$0.17 & 0.09  & 0.20 &   0.04 &  $-$1100  &700 &$-$390   & 510 &$-$110  &  180& 20    &   80\\
\object{HE2340$-$4443} & 13700& 300  & 4200  & 300 &   0.05  & 0.09  & 0.35 &   0.06 &  600    &700 &220    & 290 &180   &  220& 110   &  150\\
\object{HE2349$-$3800} & 20600& 800  & 5200  & 500 &   0.34  & 0.09  & 0.34 &   0.06 &  2300   &1000&930    & 470 &$-$200  &  180& $-$410  &  130\\
\object{HE2352$-$4010} & 14700& 2400  &3200  & 300 &   $-$0.07 & 0.08  & 0.29 &   0.05 &  $-$400   &500 &$-$310   & 280 &$-$90   &  180& $-$10   &  100\\
\object{HE2355$-$4621} & 22000& 2300  &6900  & 500 &   0.26  & 0.08  & 0.35 &   0.05 &  2900   &1000&1320   & 460 &390   &  440& $-$20   &  230\\
\noalign{\smallskip} \hline \hline
\end{tabular}
\end{center}
\begin{list}{}{}
\item[$^{\mathrm{a}}$] In units of \kms.
\item[$^{\mathrm{b}}$] 2$\sigma$ confidence level uncertainty.
\item[$^{\mathrm{c}}$] Asymmetry index defined as in Marziani et al.\ 
(\cite{M96}).
\item[$^{\mathrm{d}}$] Kurtosis parameter as in Marziani et
al.\ (\cite{M96}).
\end{list}
\end{table*}

\begin{table*}
\begin{center}
\caption{Results on \civ\ and Optical Redshift Comparison for Sources
with $z \ge 1.5$} \label{tab:civ}
    \begin{tabular}{llllll}
    \hline  \hline
    \noalign{\smallskip}
     \multicolumn{1}{c}{Source$^{\mathrm{a}}$} &
     Pop.$^{\mathrm{b}}$ &
     \multicolumn{1}{c}{$z^{\mathrm{c}}$(HE)} &
     \multicolumn{1}{c}{Shift$^{\mathrm{c}}$} &
     \multicolumn{1}{c}{Notes}\\
&&&  \multicolumn{1}{c}{[\kms]}\\  \hline
\object{HE 0005$-$2355} &  B  & 1.405      &  $-$900  \\
\object{HE 0122$-$3759} &  A  & 2.164      &  $-$3400 \\
\object{HE 0205$-$3756} &  A  & 2.404      &  $-$3000 & \hbbc\ profile of Pop.\ B; borderline \\
\object{HE 0248$-$3628} &  A  & 1.516      &  $-$2400  \\
\object{HE 2202$-$2557} &  B  & 1.5295     &  $-$600  \\
\object{HE 2352$-$4010} &  A  & 1.540      &  $-$4600 & very low W(\civ),  extreme Pop.\ A\\
\object{HE 2355$-$4621} &  B  & 2.380      &  $-$200   & excellent spectrum; \\
               &     &             &         & could be $z$(\civbc) $>$ $z$(\civnc)\\
\noalign{\smallskip}
 \noalign{\smallskip} \hline \hline
\end{tabular}
\end{center}
\begin{list}{}{}
\item[$^{\mathrm{a}}$] Of the other sources with $ z \ga 1.5$, the spectrum of \object{HE 0349$-$5249}
is not available to us. The spectrum of \object{HE 2349$-$3800} is too noisy even for rough shift
measurements.
\item[$^{\mathrm{b}}$] Classification done according to the
luminosity-dependent relationship FWHM(\Mb) $\approx 500 \times 10^{(-0.08
(M_{\rm B}+20.24))} + 3400$ \kms.
\item[$^{\mathrm{c}}$] Shift measured as $c [z(\hbox{\civ)}-z_{\rm opt}] / [1. + z_{\rm opt}]  $, where
$z_{\rm opt}$ is as reported in Table~\ref{tab:obs}, and $z$(\civ)  refers to a
measurement of the upper half of the profile by Gaussian fitting. No
attempt was made to deconvolve \civbc\ and \civnc.
\end{list}
\end{table*}

\begin{table*}
\begin{center}
\caption{Results of Luminosity Correlation Analysis} \label{tab:corr}
    \begin{tabular}{lllllllllll}
    \hline  \hline
    \noalign{\smallskip}
     \multicolumn{1}{c}{Parameters$^{\mathrm{a}}$} & Sample$^{\mathrm{b}}$ &
     & \multicolumn{3}{c}{Spearman's $\rho^{\mathrm{c}}$}\\
 &&& \multicolumn{1}{c}{All} & \multicolumn{1}{c}{RQ$^{\mathrm{d}}$} &
\multicolumn{1}{c}{RL$^{\mathrm{d}}$}
\\
 \multicolumn{1}{c}{(1)}      & \multicolumn{1}{c}{(2)} & &
\multicolumn{1}{c}{(3)}
     & \multicolumn{1}{c}{(4)}   &  \multicolumn{1}{c}{(5)}
     \\    \hline
\noalign{\smallskip}
FWHM & M03 &&  $-$0.159 (0.020) & $-$0.066 (0.418)  & 0.092 (0.463)  \\
FWHM & M03+VLT/ISAAC && $-$0.165 (0.012) & $-$0.105 (0.183) &  0.087 (0.475)   \\
FWHM & PG+VLT/ISAAC && $-$0.330 (0.001) & $-$0.261 (0.016) &  0.155 (0.561)   \\
\\
\rfe\ UL   & M03 && 0.222 (0.001) &  0.087 (0.290) & 0.159 (0.202)   \\
\rfe\ UL   & M03+VLT/ISAAC && 0.160 (0.015) & 0.038 (0.623) & 0.079 (0.515)   \\
\rfe\ UL   & PG+VLT/ISAAC && 0.069 (0.489) & $-$0.011 (0.922) & $-$0.017 (0.951)   \\
\\
W(\hbbc) & M03 && $-$0.073 (0.288) & $-$0.136 (0.096) & 0.088 (0.481) \\
W(\hbbc) & M03+VLT/ISAAC && $-$0.056 (0.734) & 0.088 (0.2695) & 0.029 (0.810) \\
W(\hbbc) & PG+VLT/ISAAC && 0.169 (0.090) & 0.090 (0.4015) & 0.689 (0.010) \\

\\
W(\feiiq) UL & M03    && 0.199 (0.004) & $-$0.019 (0.8184) & 0.156 (0.213)  \\
W(\feiiq) UL & M03+VLT/ISAAC    && 0.184 (0.005) & $-$0.001 (0.990)  &  0.174 (0.157)   \\
W(\feiiq) & PG+VLT/ISAAC && 0.130 (0.191) & $-$0.021 (0.847)& 0.439 (0.100) \\
 \\
W(\oiii) & M03 &&  0.300 ($< 10^{-4}$) &  0.464 ($< 10^{-4}$) & 0.317 (0.012) \\
W(\oiii) & M03+VLT/ISAAC && 0.375 ($< 10^{-4}$) &  0.541 ($< 10^{-4}$) & 0.319
(0.009) \\
W(\oiii) & PG+VLT/ISAAC  && 0.332 (0.001)  & 0.423 (0.0001) & 0.300 (0.261) \\
 \noalign{\smallskip} \hline \hline
\end{tabular}
\end{center}
\begin{list}{}{}
\item[$^{\mathrm{a}}$] Correlated against \Mb\ ($q_0 = 0$; $H_0 = 75$ \kms\ Mpc$^{-1}$).
\item[$^{\mathrm{b}}$] M03: 215 objects in Marziani et al.\ (\cite{M03a}); 
VLT/ISAAC: 17  observations of this paper.
\item[$^{\mathrm{c}}$] Computed with the assumption that \feiiq\ upper limits
(UL) are censored data. If no UL are present, $\rho$\ is equal to the
Spearman's rank correlation coefficient $r$. In parenthesis, we report the
probability that a correlation is not present.
\item[$^{\mathrm{d}}$] M03: 65 RL, 150 RQ; M03+VLT/ISAAC: 69 RL, 162 RQ; PG+VLT/ISAAC: 102 (85 PG) sources;
87 RQ and 15 RL.

\end{list}
\end{table*}

\newpage

%\listofobjects

\end{document}